\documentclass[preprint2,numberedappendix,iop]{emulateapj-rtx4}

\usepackage{graphicx,graphics,amsmath}
\usepackage{natbib}
\usepackage{bm,url}
\usepackage{times}
\usepackage{amssymb}
\usepackage{color}
\usepackage{float}
\graphicspath{{./fig/}{./png/}}
\newcommand{\EQ}{\begin{equation}}
\newcommand{\EN}{\end{equation}}
\newcommand{\EQA}{\begin{eqnarray}}
\newcommand{\ENA}{\end{eqnarray}}

\newcommand{\mps}{m~s$^{-1}$}

\newcommand{\cmss}{cm$^2$~s$^{-1}$}
\newcommand{\vp}{$\gamma_{r0}$}
\def\mc{meridional circulation}
\def\mf{meridional flow}

\def\bl{Babcock--Leighton}
\def\ftdm{flux transport dynamo model}
\def\ftdms{flux transport dynamo models}

\newcommand{\rhat}{\hat{\mbox{\boldmath $r$}} {}}
\newcommand{\thetahat}{\hat{\mbox{\boldmath $\theta$}} {}}
\newcommand{\phihat}{\hat{\mbox{\boldmath $\phi$}} {}}
\newcommand{\etas}{\eta_{\mathrm{surf}}}
\newcommand{\etab}{\eta_{\mathrm{CZ}}}
\newcommand{\etaRZ}{\eta_{\mathrm{RZ}}}

\newcommand{\mean}[1]{\overline{#1}}

\newcommand{\Eq}[1]{Equation~(\ref{#1})}
\newcommand{\Eqs}[2]{equations~(\ref{#1}) and~(\ref{#2})}

\newcommand{\Sec}[1]{\S\ref{#1}}

\newcommand{\Fig}[1]{Figure~\ref{#1}}

\newcommand{\Figs}[2]{Figures~\ref{#1} and \ref{#2}}

\usepackage{color}
\def\blue{\textcolor{black}}

\shorttitle{Babcock-Leighton solar dynamo}
\shortauthors{Karak \& Cameron}

\begin{document}

\title{Babcock-Leighton solar dynamo: the role of downward pumping and the equatorward propagation of activity}
\author{Bidya Binay Karak$^{1,2}$ and Robert Cameron$^{1}$}
\affil{$^1$Max-Planck-Institut f\"ur Sonnensystemforschung, Justus-von-Liebig-Weg 3, D-37077 G\"ottingen, Germany\\
$^1$High Altitude Observatory, National Center for Atmospheric Research, 3080 Center Green Dr., Boulder, CO 80301, USA}
%Keywords: Key words: dynamo – Sun: activity – Sun: magnetic fields
%%Keywords: Sun: dynamo, Sun: magnetic fields, Sun: activity, (Sun:) sunspots, Sun: interior, Sun: rotation  
 \begin{abstract}

The key elements of the Babcock-Leighton dynamos
are the generation of poloidal field 
through decay and dispersal of tilted bipolar active regions and the generation of toroidal field 
through the observed differential rotation.
These models are traditionally known as flux transport dynamo models
as the equatorward propagations of the butterfly wings in these models are 
produced due to an equatorward flow at the bottom of the convection zone.
Here we investigate the role of downward magnetic pumping near the surface 
using a kinematic Babcock-Leighton model. 
We find that the pumping causes the poloidal field to become predominately radial in
the near-surface shear layer, which allows the negative radial shear to
effectively act on the radial field to produce a toroidal field.
We observe a clear equatorward migration of the
toroidal field at low latitudes as a consequence of the dynamo wave
even when there is no meridional flow in the deep convection zone. 
Both the dynamo wave and the flux transport 
type solutions are thus able to reproduce some of 
the observed features of the solar cycle, including the 11-year periodicity. 
The main difference between the two types of 
solutions is the strength of the Babcock-Leighton source 
required to produce the dynamo action.
A second consequence of the magnetic pumping is that it suppresses the diffusion of 
fields through the surface, which helps to allow an 11-year cycle at
(moderately) larger values of magnetic diffusivity than have previously been used.

  \end{abstract}

  \email{bkarak@ucar.edu}
  \maketitle
%____________________________________________________________

\section{Introduction}
In most models of the solar dynamo, there are two key processes: 
the generation of toroidal magnetic field by winding up of
poloidal field by differential rotation ($\Omega$ effect) and the generation of 
poloidal magnetic field by flows associated with small-scale
motions, influenced by the Coriolis force, acting on the toroidal field ($\alpha$ effect);
see \citet{C14} for a review.
Recently we have strong observational supports that the dynamo is
of the Babcock-Leighton type, 
where the poloidal field is produced from the decay and dispersal of tilted 
active regions near the surface \citep[e.g.,][]{Das10,KO11,Cameron15}.
Although these papers constrain the nature of dynamo mechanism operating in the Sun, 
they do not address the questions of why the butterfly wings (the latitudes at which sunspots emerge) 
drift equatorward, and why the dynamo period is about 11~years. 
%We begin to address these questions in this paper.

We are not the first to address these questions. 
For the equatorward migration, two main possible mechanisms are 
available in the literature.
The first mechanism to appear was that of \cite{Pa55}, 
and it corresponds to a dynamo wave, where the propagation is on surfaces
of constant rotation \citep{Yo75}. 
Hence, the radial differential rotation gives rise to a latitudinal migration.
This is known as the Parker-Yoshimura sign rule. 
According to this rule, the direction of propagation
(poleward or equatorward) depends on the sign of the $\alpha$ effect 
(the sense of the helical motion), 
and the sign of the radial shear. 
To have equatorward propagation in the Sun 
where $\alpha$ is positive in the northern hemisphere, differential rotation must increase inwards \citep{Yo75}.

From 1955 to the mid-1980s, this dynamo-wave model was widely accepted.
However, a decline in the popularity of this model followed from a combination of the suggestion
by \cite{Parker75} that, in order to maintain organized toroidal flux over many years,
the field must be stored at the base of the solar convection zone (SCZ), 
%and the discovery using helioseismology that the differential rotation in the tachocline
and the discovery using helioseismology that the differential rotation in the 
\blue{
low-latitude
}
tachocline
decreases inwards, which would lead to poleward propagation
of the butterfly wings -- contrary to what is observed.
One possibility to obtain the equatorward migration with the field still being at the
  base of CZ, where the radial shear is positive, 
is if
the $\alpha$ is negative (in the northern hemisphere) there,
as is observed in some three-dimensional convection simulations
\citep[e.g.,][]{Br90,ABMT15, War16}, 

The question of how to obtain the equatorward migration of the activity belt outside the
dynamo-wave framework led to the identification of a second possible mechanism, 
the flux transport dynamo (FTD) model \citep{WSN91,CSD95,Dur95}.
In this model, it is assumed that an equatorward velocity near the base of the CZ
overpowers the poleward propagating dynamo wave and leads to a net equatorward migration in 
the toroidal flux that produces sunspot eruptions.

However, more recently three-dimensional convective dynamo simulations have shown, 
contrary to the suggestion of \cite{Parker75},
that substantial organized toroidal magnetic field can be stored 
in the CZ for periods of years 
\citep{Brown10}. 
Furthermore, storage in the
tachocline has been shown to be problematic \citep{Weber15}, 
and concerns have been raised about the ability to extract energy
from the radial shear in the tachocline 
\citep{VB09,Spruit11}. 
This has lead to an increasing support for the dynamo wave explanation for the
equatorward migration 
with the near-surface shear layer (NSSL),
where the differential rotation increases inwards
\citep[e.g.,][]{Br05,KKT06,Brandenburg09,PK11}. 

Returning to the other question---why the dynamo period is 11 years---sheds light
on the two possibilities. In a simple oscillatory $\alpha\Omega$ dynamo, the cycle period
is inversely related to the turbulent diffusivity $\eta$ as well as dependence on
$\alpha$ and 
shear. 
This leads to a short cycle period of 2--3 years 
\cite[using a value of $\eta\sim10^{12-13}$~\cmss --- an estimation 
based on the mixing length theory;][]{kohler73,KC12}. 
A similar difficulty also appears in the \bl\ type FTD models.
Therefore FTD models have used much lower values of $\eta$ than the mixing length estimates, 
so that the cycle period is predominately determined by the speed of the
meridional flow near the bottom \citep{DC99,Kar10}.

One physical justification for the use of this weaker value of $\eta$ in dynamo
models was that the magnetic quenching could reduce $\eta$ significantly from its mixing
length value \citep{RKKS94,GDG09,MNM11}. Indeed, there is evidence of magnetic quenching in various
numerical simulations, although most of the simulations at present are 
performed in parameters regimes far from the SCZ \citep[e.g.,][]{KB09,Kar14b}.
The latter authors, however, found substantial quenching only when the mean magnetic field is
substantially greater than the equipartition field strength (based on the energy in the
convective motions). In particular, a reduction of $\eta$ by two orders of magnitude
(from $10^{13}$~\cmss\ to $10^{11}$~\cmss)
requires the mean magnetic field to be larger than the equipartition value by 
more than two orders of magnitude 
\cite[also see][who support this result]{SCD16}.
Such field strengths are not plausible. Another possibility for weaker diffusion
is that changes in the mean flows (e.g., the inflow observed around active regions) 
could play a role \citep{CS16}.
In any case, exploring the range of values of $\eta$ for which the \bl\ dynamos have
equatorial propagation and the 11-year period is a useful exercise.

In this article, we construct a Babcock-Leighton dynamo model
and explore how the radial pumping affects equatorward migration and the cycle period
in both FTD and dynamo wave frameworks.
%An important ingredient of both models is the differential rotation, 
%and for this we consider two different profiles.
For both models, differential rotation is an important ingredient
for which we consider two different profiles.
One closely corresponds to the available helioseismic data, 
while the other is an analytic approximation that includes the observed NSSL but
neglects the radial variation in most of the CZ---from 
top of the tachocline to the bottom of the NSSL.
The motivation for the latter profile is that
%it gives the maximal toroidal flux generation from magnetic flux threading the surface
it gives a maximal toroidal flux generated from magnetic flux threading the surface
by the radial shear because in helioseismic profile, the radial shear has the opposite sign above and below
the bottom of the NSSL.
We show that for both choices of differential rotation 
the NSSL where the rotation rate increases inward plays an
important role in producing the equatorward migration of sunspots in addition to the
migration caused by the equatorward meridional flow near the bottom of the CZ.
We demonstrate that a radially downward magnetic pumping makes the magnetic field 
more 
radial 
in the NSSL, which allows the (negative) radial shear to stretch the radial field to 
%make it toroidal. 
\blue{
generate a toroidal field. 
}
This helps to produce an equatorward migration of toroidal field where the (Babcock-Leighton)
$\alpha$ is non-zero.

Moreover, the magnetic pumping inhibits the diffusion of toroidal field through the photosphere
that helps to excite dynamo with a correct period at a considerably higher value of $\eta$ than it was possible
in earlier. This role of the pumping is complementary to that
studied by \citet{Ca12} who further demonstrated that a downward pumping is crucial to
match the results of FTDs with the surface flux transport model and observations.

\section{Model}
In our model, we study the axisymmetric large-scale magnetic field 
in the kinematic regime.
Therefore the magnetic field can be written as
\begin{equation}
{\bf B} = {\bf B_p} + {\bf B_\phi} =\nabla \times [ A(r, \theta, t) \phihat] + B (r, \theta, t) \phihat,
\end{equation}
where ${\bf B_p} = \nabla \times [ A \phihat]$ 
is the poloidal component of the magnetic field and
$B$ is the toroidal component.
Similarly the velocity field can be written as
\begin{equation}
{\bf v} = {\bf v_p} + {\bf v_\phi}\\
= v_r(r,\theta) \rhat + v_\theta(r,\theta) \thetahat + r \sin \theta \Omega(r,\theta) \phihat,
\label{eqv}
\end{equation}
where $v_r$ and $v_\theta$ correspond to the meridional circulation and
$\Omega$ is the angular frequency.
Then, the evolution equations of $A$ and $B$ in the \ftdm\
become followings.
\begin{equation}
\frac{\partial A}{\partial t} + \frac{1}{s}({\bf v_p}.\nabla)(s A)
= \eta \left( \nabla^2 - \frac{1}{s^2} \right) A + S(r, \theta; B),
\label{eqpol}
\end{equation}
\begin{eqnarray}
\frac{\partial B}{\partial t}
+ \frac{1}{r} \left[ \frac{\partial}{\partial r}
(r v_r B) + \frac{\partial}{\partial \theta}(v_{\theta} B) \right]
= \eta \left( \nabla^2 - \frac{1}{s^2} \right) B \nonumber \\
+ s({\bf B_p}.{\bf \nabla})\Omega + \frac{1}{r}\frac{d\eta}{dr}\frac{\partial{(rB)}}{\partial{r}},~~~~~~~~~~~~~~~~~~~~~~~~~~~~~
\label{eqtor}
\end{eqnarray}\\
where $s = r \sin \theta$.

\begin{figure}
\centering
\includegraphics[width=1.05\columnwidth]{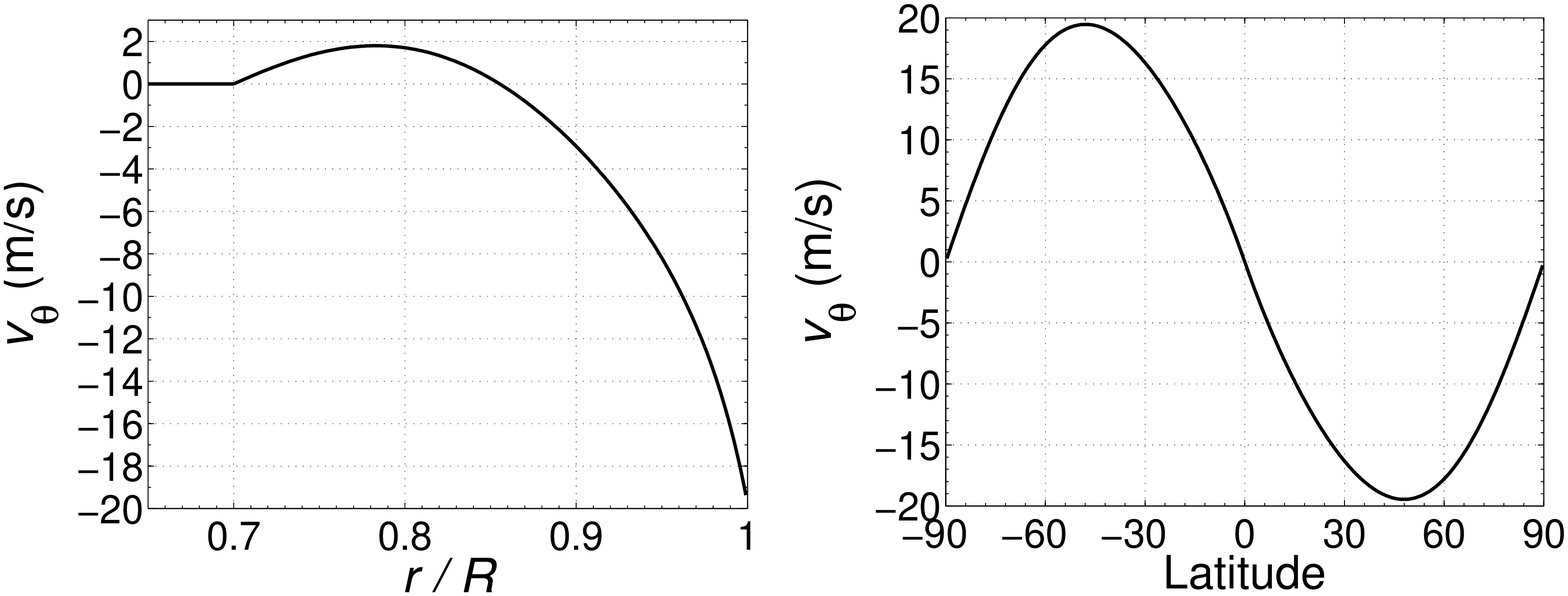}
\caption{Radial and latitudinal dependences of the latitudinal component of the
meridional flow $v_\theta$ at $45^\circ$ latitude (left panel) and
at the surface (right), respectively.
}
\label{fig:mc}
\end{figure}

For the meridional circulation we define a stream function $\psi$ such that 
$\rho {\bf v_p} = \nabla \times [\psi (r, \theta) {\bf e}_{\phi}],$
where 
%\begin{equation}
$\rho = C \left( \frac{R}{r} - 0.95 \right)^{3/2}, \label{rho}$
%\end{equation}
and
\begin{eqnarray}
\label{eq:psi}
\psi r \sin \theta = \psi_0 (r - R_p) \sin \left[ \frac{\pi (r - R_p)}{(R -R_p)} \right]\{ 1 - e^{- \beta_1 \theta^{\epsilon}}\}\nonumber \\
\times\{1 - e^{\beta_2 (\theta - \pi/2)} \} e^{-((r -r_0)/\Gamma)^2} ~~~~
 \label{eqmc}
\end{eqnarray}\\
with
$\beta_1 = 1.5, \beta_2 = 1.3$, $\epsilon = 2.0000001$, $r_0 = 0.45R/3.5$, $\Gamma =
3.47 \times 10^{8}$ m, and $R_p = 0.7 R$.
The value of $\psi_0/C$ is chosen in such a way that the amplitude of the meridional circulation
at mid-latitudes $v_0$ becomes $20$~\mps.
Our profile is very similar to many previous publications, particularly
\citet{HKC14} (see their Eqs.~6-8) except here we assume that the \mf\ smoothly goes to zero at $0.7R$. 
The resulting variation is shown in \Fig{fig:mc}.
We note that our results are not very sensitive to the detailed flow structure 
%and the direction as long as we have a reasonable amount of poleward flow near the surface.
\blue{
in the CZ as long as we have a reasonable amount of poleward flow near the surface.
}
Towards the end of this article (\Sec{sec:mc1}), we shall show that we get 
the correct magnetic cycle 
even with a shallow meridional circulation residing only in the upper $0.8R$.

In addition to the large-scale \mc, we add a magnetic pumping, the $\gamma$ effect
in our model. This $\gamma$ appears as an advective term in the mean-field 
induction equation. 
Unlike the large-scale circulation, $\gamma$ is not divergenceless.
%While there may be some uncertainties
%about the existence and the nature of the latitudinal pumping in the SCZ, the
%radial pumping is ubiquitous.
%Many convection simulations predict a downward magnetic pumping 
%$\gamma_r$ with magnitude of several meters per second \citep[e.g.,][]{Kap06,Kar14b,ABMT15,War16}. 
%Although we do not know the radial variation of $\gamma_r$, we assume that it is only significant
%in the top $10\%$ of the Sun. 
Theoretical analysis and local magneto-convection simulations predict 
a downward magnetic pumping in the SCZ
\citep{DY74, KR80, PS93, Tob98, KKB09, Kar14b}.
%The recent global convection simulations in stellar CZs \citep[e.g.,][]{ABMT15,War16} also 
The recent 
\blue{
rapidly rotating convection simulations 
}
in stellar CZs \citep[e.g.,][]{ABMT15,War16} also 
find magnetic pumping, although in some cases, it is radially outward
at some latitudes. 
%These simulations, however, are not the end of the story
These simulations, however, may not represent the actual picture
because they do not capture the realistic physics of the solar surface convection and even 
they do not extend to the top surface, where a strong pumping can be
inferred from the success of the surface flux transport model \citep{Ca12}.
\citep[See also][who demonstrated the importance of downward pumping 
in simulating the solar cycle.]{GD08,KN12,JCSI13}
Hence, in our study we assume that it is downward and only significant
in the top $10\%$ of the Sun.
This assumption is supported by a number of facts that 
the upper layer of the Sun is highly unstable to convection while the deeper 
layer is only weakly unstable \citep{spr97}, convection is very weak below the 
surface layer \citep{Hanasoge12}, and a huge density stratification near the surface.

%Furthermore, \citet{Ca12} previously demonstrated that 
%the magnetic pumping in the near-surface shear layer is crucial to match 
%the results of a FTD with that
%of a surface flux transport model.

Therefore in \Eq{eqv}, we replace $v_r$ by  
$v_r+ \gamma_r$ where
\begin{equation}
\gamma_r(r) = - \frac{\gamma_{r0}}{2}\left[1 + \mathrm{erf} \left(\frac{r - 0.9R}{0.02R}\right) \right].
\label{eqpump}
\end{equation}
Due to the lack of knowledge of the exact latitudinal variation of $\gamma_r$, 
we take it to be only a function of radius.
This is indeed a good choice based on the high Rossby number of the surface convection.
The value of \vp\ is not directly constrained by observations 
and we take it as a free parameter.

\begin{figure}
\centering
\includegraphics[width=0.8\columnwidth]{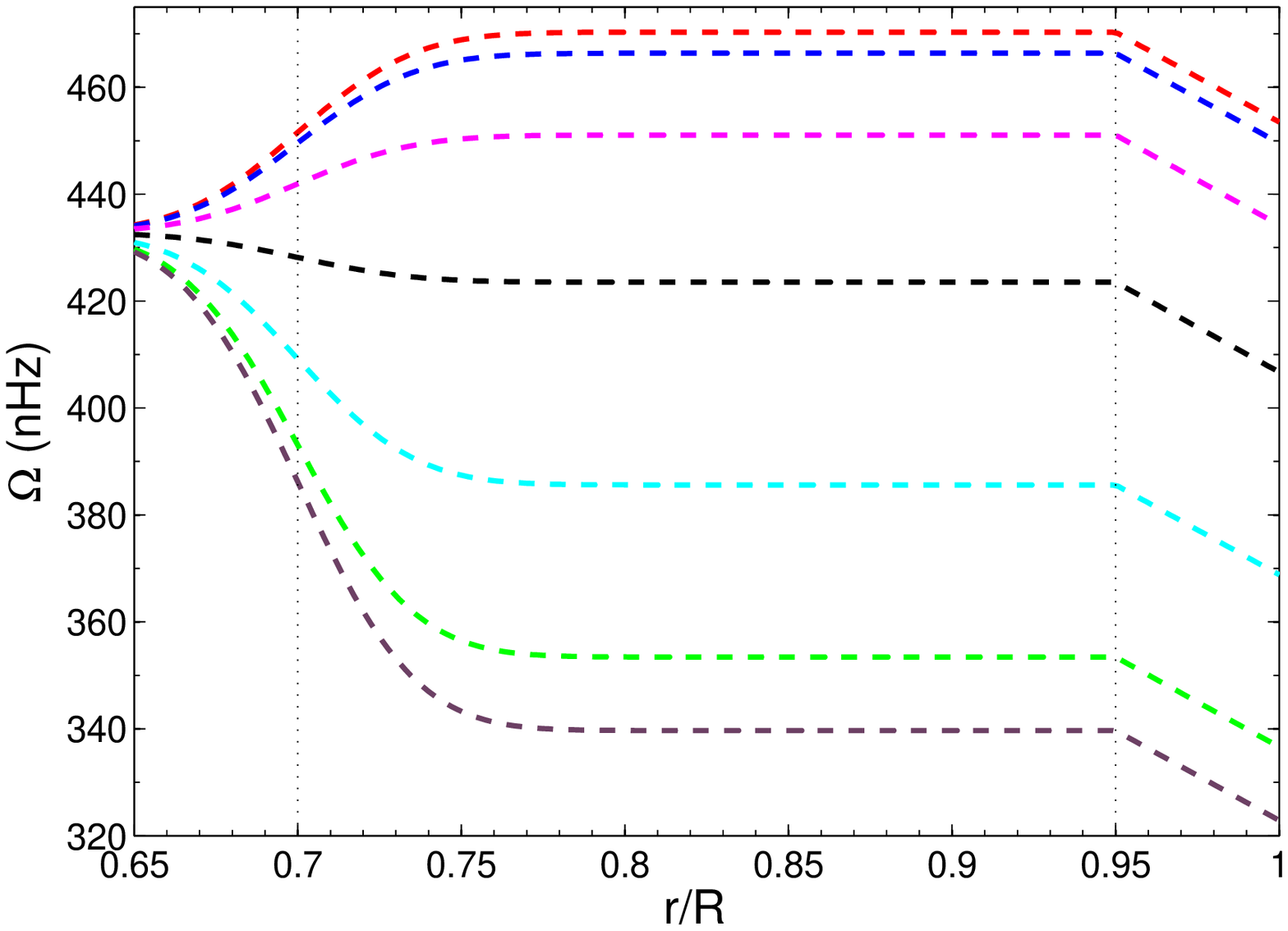}
\caption{Radial variations of the angular frequencies at 
0 (red), 15, 30 (pink), 45, 60, 75 (green) and 90 degree 
latitudes based on \Eq{eq:oma}.
}
\label{omgan}
\end{figure}

\begin{figure}
\centering
\hspace*{-0.8cm}
\includegraphics[width=1.20\columnwidth]{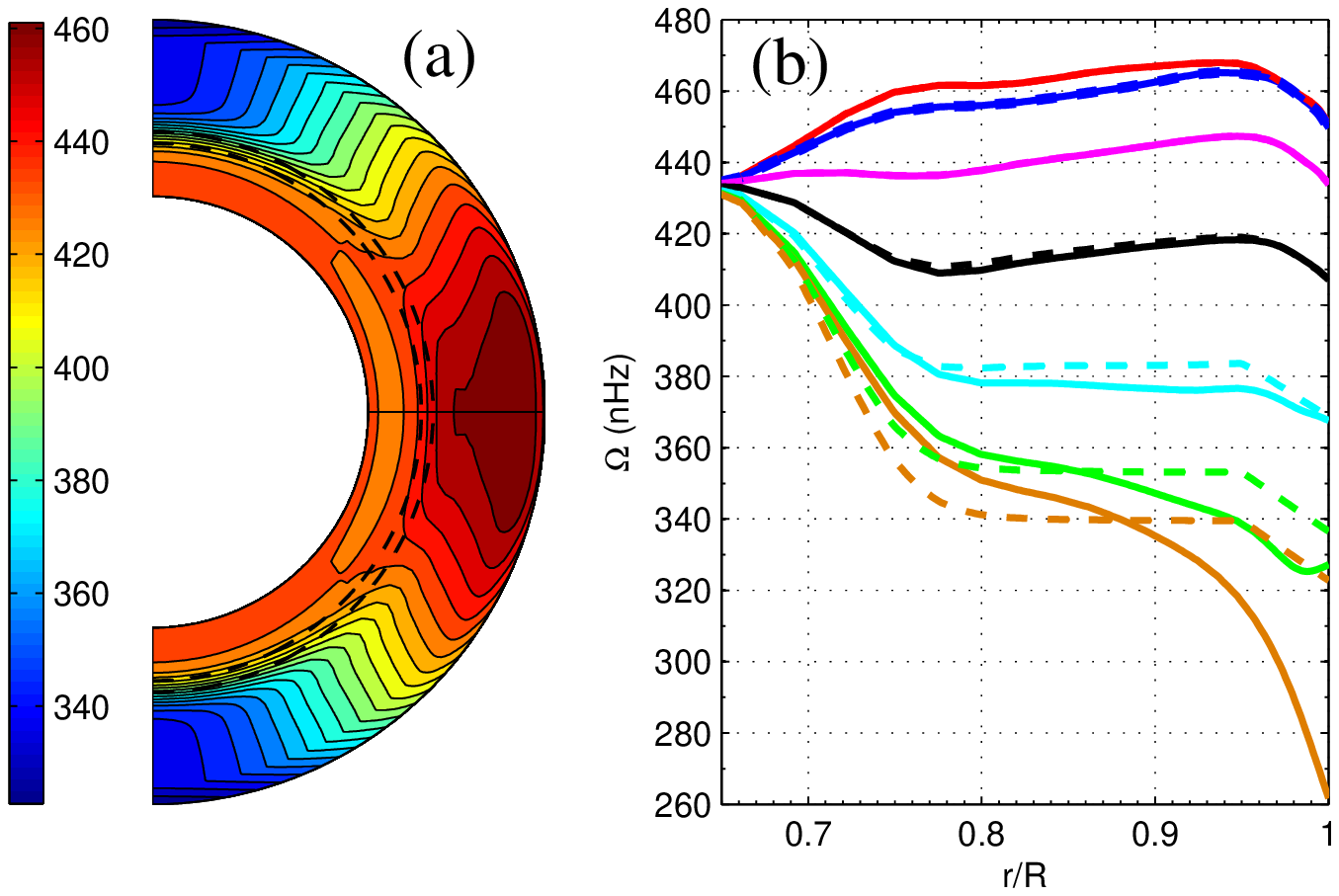}
\caption{(a): Rotational frequencies of Sun (in nHz) as obtained from helioseismic data 
combined with the analytical profile for higher latitudes. 
(b): comparison of helioseismic data (solid lines) and our composite (dashed) 
profile at 0 (red), 15, 30 (pink), 45, 60, 75 (green) and 90 
degree latitudes.
}
\label{omgob}
\end{figure}

For angular frequency we take two different profiles. First, we 
choose a simplified analytical form given by
\begin{equation}
\Omega(r,\theta) = \Omega_\mathrm{RZ} + \frac{\Omega_\mathrm{CZ} - \Omega_\mathrm{RZ}}{2}\left[1 + \mathrm{erf} \left(\frac{r - 0.7R}{0.04R}\right) \right].
\label{eq:oma}
\end{equation}
Here $\Omega_\mathrm{RZ}/2\pi = 432.8$~nHz, and
$\Omega_\mathrm{CZ}/2\pi = \Omega_1$ for $r < 0.95R$, while 
$\Omega_\mathrm{CZ}/2\pi = \Omega_1 - a_\mathrm{s}(r-0.95R)/0.05R$ for $r \ge 0.95R$,
%where $\Omega_1 = 453.5 - 62.79\cos^2\theta -  67.87\cos^4\theta + 16.808 $~nHz
where $\Omega_1 = 470.308 - 62.79\cos^2\theta -  67.87\cos^4\theta$~nHz
and $a_\mathrm{s}=16.808$~nHz.
This type of simple profile, except the near-surface shear layer, is
used in many previous \ftdms\ 
\citep[except,][]{Dik02,GD08}.
The radial variation of this $\Omega$ at different latitudes is shown
in \Fig{omgan}.
Later in \Sec{sec:Omobs} we shall use the observed rotation profile from
helioseismology, kindly provided by J.\ Schou.
%BBK: Which data it is? Shall we write it?
However, these measurements are not reliable in high latitudes. Therefore
we construct a composite profile using the above analytical profile for high
latitudes where the observed data is less reliable.
The composite $\Omega$ to be used in our dynamo model is shown 
in \Fig{omgob}.

\begin{figure}
\centering
\includegraphics[width=0.65\columnwidth]{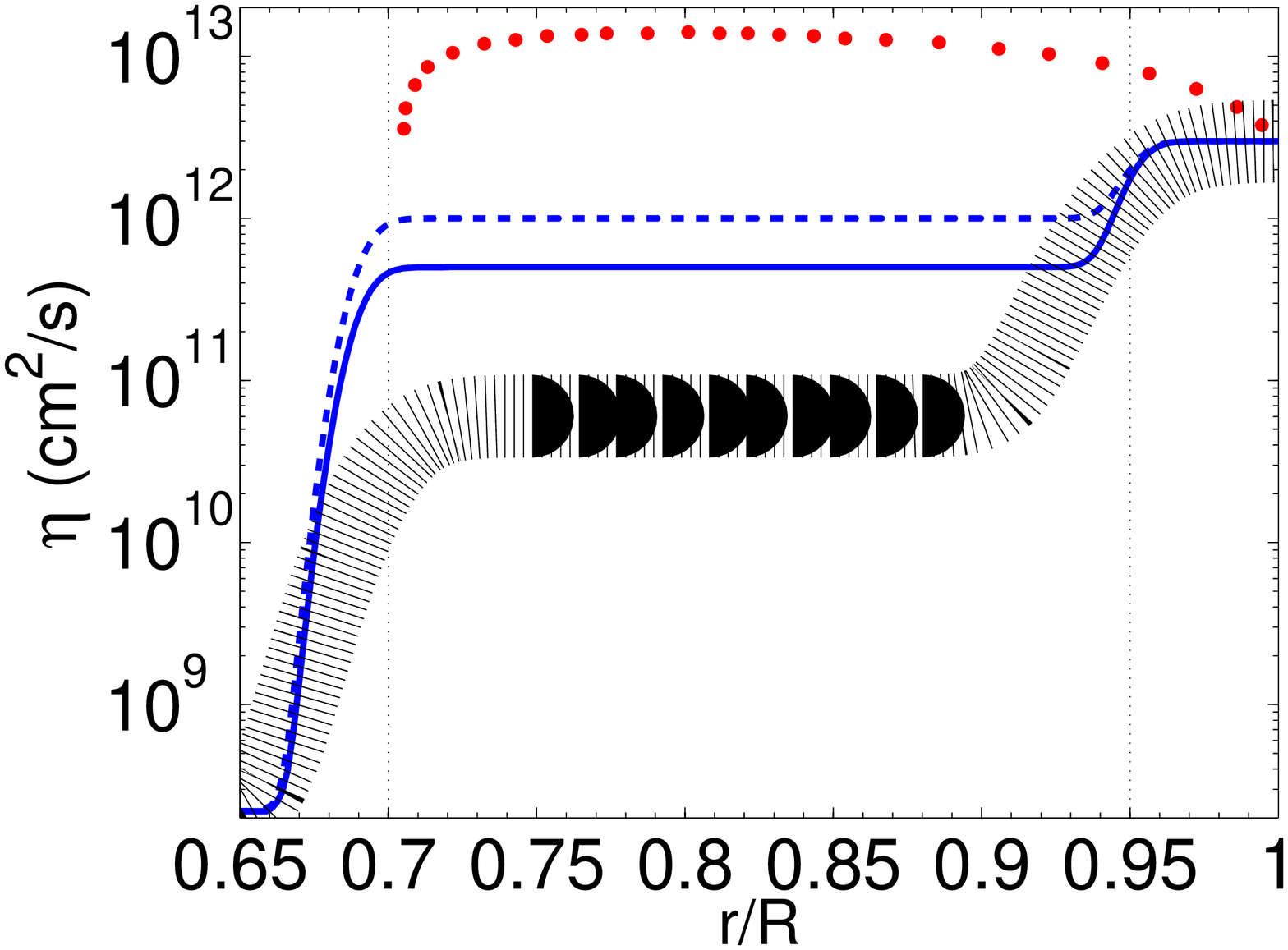}
\caption{Radial variations of the turbulent diffusivity (solid line)
and its allowed upper value (dashed line) in our model. 
Lower shaded region and upper red points 
represent typical values used in most previous FTD models and 
the mixing-length theory.
}
\label{fig:eta}
\end{figure}

For the turbulent magnetic diffusivity, we choose the following radial dependent profile
motivated by previous models \citep[e.g.,][]{Dik02}:
\begin{eqnarray}
\eta(r) = \etaRZ + \frac{\etab-\etaRZ}{2}\left[1 + \mathrm{erf} \left(\frac{r - 0.69R}{0.01R}\right) \right]+ \nonumber  \\
\frac{\etas-\etab-\etaRZ}{2}\left[1 + \mathrm{erf} \left(\frac{r - 0.95R}{0.01R}\right) \right],
\label{eqeta}
\end{eqnarray}\\
where
$\etaRZ = 2.2 \times 10^8$~cm$^2$~s$^{-1}$, $\etas = 3 \times 10^{12}$~cm$^2$~s$^{-1}$.
While the diffusivity near the surface is constrained by observations as well as 
by the surface flux transport model \citep[e.g.,][]{KHH95,LCC15}, 
the diffusivity in the deeper CZ is poorly
known. For most of the simulations, we choose $\etab = 5 \times 10^{11}$~cm$^2$~s$^{-1}$; 
however in some simulations we vary it up to $10 \times 10^{11}$~cm$^2$~s$^{-1}$.
The radial dependence of $\eta$ is shown in \Fig{fig:eta}.

Finally, the source term $S(r, \theta; B)$, which captures the Babcock--Leighton 
mechanism for the generation of poloidal field near the solar surface 
from the decay of tilted bipolar active regions, has the following form:
\begin{equation}
 S(r, \theta; B) = \alpha^{\rm BL}(r,\theta) \mean B(\theta).
\label{BLsource}
\end{equation}
Here,
\begin{equation}
\alpha^{\rm BL}(r,\theta) = \alpha_0 f_{\alpha}(\theta) \frac{1}{2}
\left[1+\mathrm{erf}\left(\frac{r-0.95R}{0.01R}\right)\right],
\label{eqalpha}
\end{equation}
where
$\alpha_0$ determines the strength of the \bl\ $\alpha$
which is set to a value for which we get a nearly stable solution, 
and
\begin{equation}
f_{\alpha}(\theta) = \cos\theta \sin^n\theta.
\label{eqlal}
\end{equation}
The radial dependence in \Eq{eqalpha} ensures that the \bl\ process only operates near the surface above $0.95R$. 
The latitudinal dependence of $f_{\alpha}(\theta)$ 
is chosen in such a way that $S(r, \theta; B)$ is negligible above $\pm30^\circ$
latitudes where we do not observe much sunspots.
This sets the value of $n$ to 12.
But in a few cases, we change it to two.
We note that previous dynamo modelers also artificially suppress the 
production of poloidal field above $30^\circ$ latitudes \citep[e.g.,][]{KKC14,MNM11,MT16}.

In some simulations discussed below, we take 
$\mean B(\theta)$ as the mean (area normalized) toroidal field over the whole CZ:
$\int_{0.7R}^{R} B(r,\theta) r dr / \int_{0.7R}^{R} r dr$.
For comparison with traditional FTD simulations, in some cases, 
we also take $\mean B(\theta)$ as the radial-averaged toroidal field
over the tachocline from $ 0.7 R$ to $0.71 R$.
Which choice  we make for $\mean B(\theta)$ will be given below in the description
of individual numerical experiments we have performed. 
We note that in our $\alpha$ (\Eq{BLsource}), we do not 
include any nonlinearity to saturate the poloidal field production,
which is common in any kinematic dynamo model \citep{Kar14a}.

The computational domain of our model is 
$0.55R \le r \le R$ and $0 \le \theta \le \pi$.
For the boundary conditions, we use 
$A, B=0$ at poles and at the bottom, while
$\frac{\partial}{\partial r}(rA), B = 0$ (radial field condition) at the top.
Our model is formulated using the code SURYA, developed
by A.\ R.\ Choudhuri and his colleagues at Indian Institute of Science \citep{CNC04}.

%:::::::::Results:::::::::::::::::::::::::::::::::::::::::::
\section{Results with simplified rotation profile}
\label{sec:Oman}
To study the evolution of magnetic fields we solve \Eqs{eqpol}{eqtor} by prescribing 
all ingredients described above.
For the first set of numerical experiments, we used the simple analytical profile for the differential rotation as 
shown in \Fig{omgan}. The advantage of using this profile is that it allows us to more easily understand the
dynamics of the system. It also corresponds to the solution that maximally exploits the near-surface shear layer
and hence provides the limit where the dynamo wave generated toroidal flux is maximal.
When the source term in \Eq{eqpol} is non-zero, we expect oscillatory dynamo solution provided
the diffusive decay of the field is less than its generation. 

\subsection{Evolution of magnetic fields for $\alpha^{\rm BL} = 0$}
\label{sec:decay}
\begin{figure}
\centering
\hspace{-0.8cm}
\includegraphics[width=0.40\columnwidth]{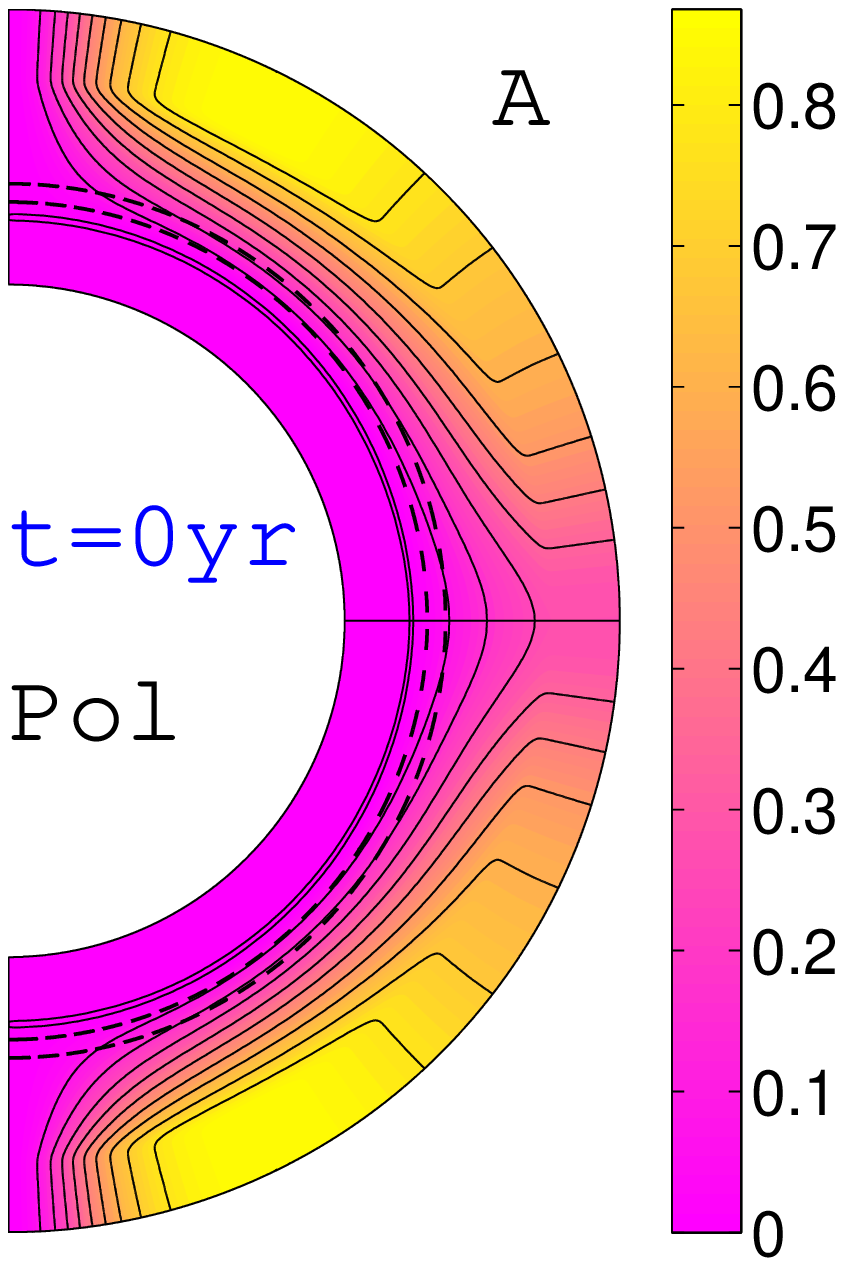}
\hspace{+0.5cm}
\includegraphics[width=0.30\columnwidth]{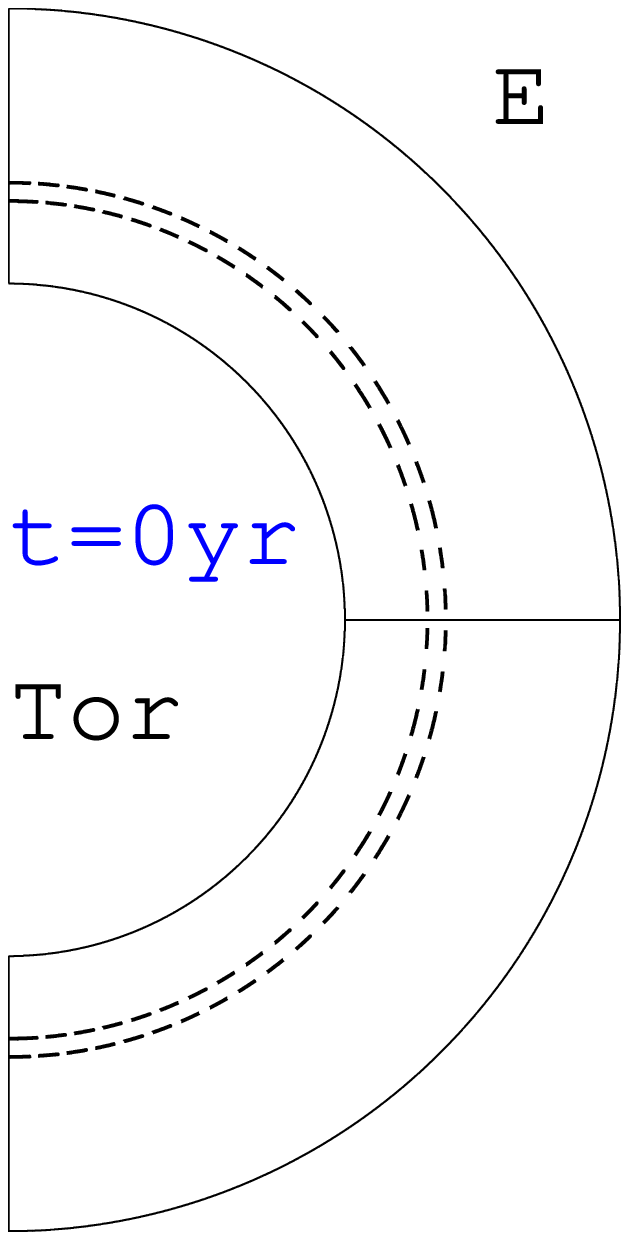}
\includegraphics[width=0.40\columnwidth]{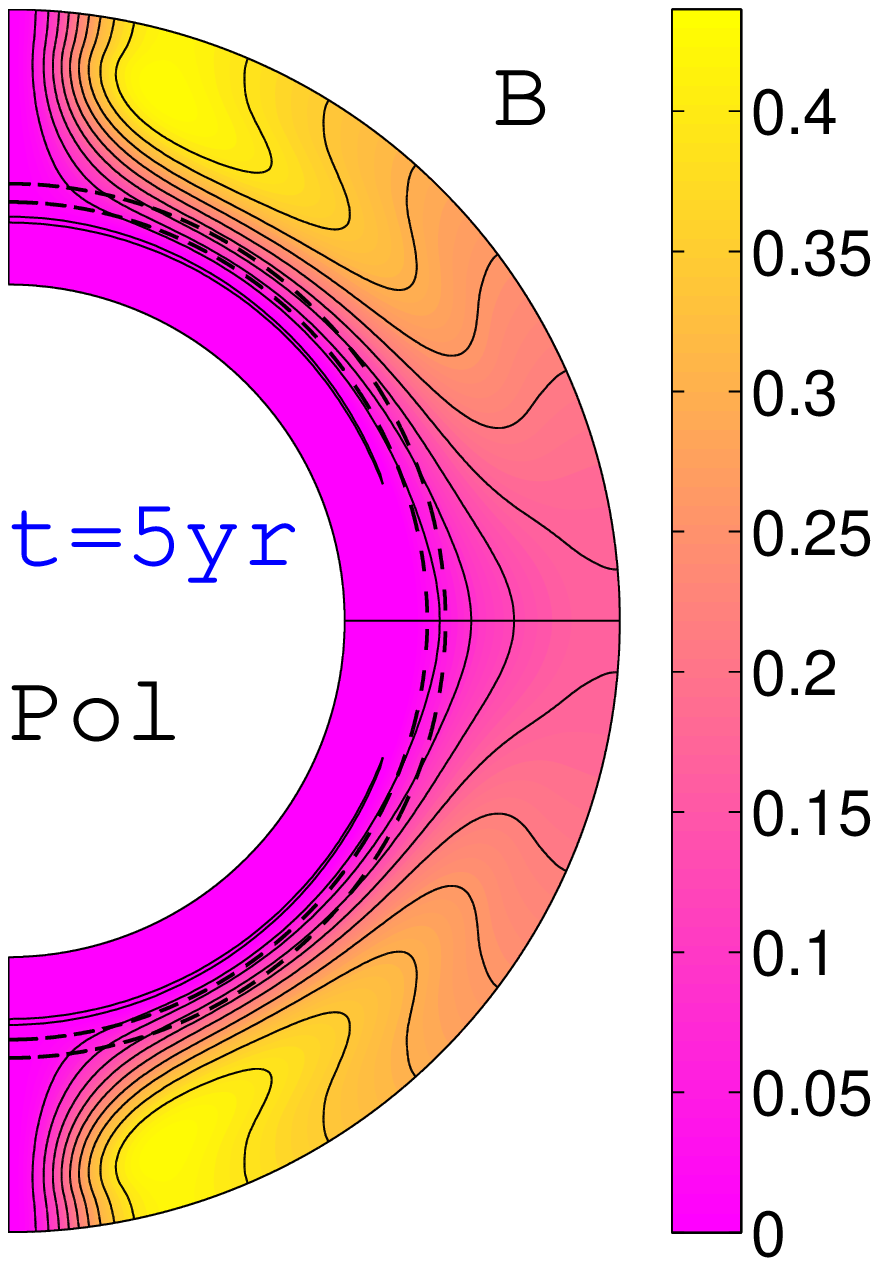}
\hspace{+0.5cm}
\includegraphics[width=0.39\columnwidth]{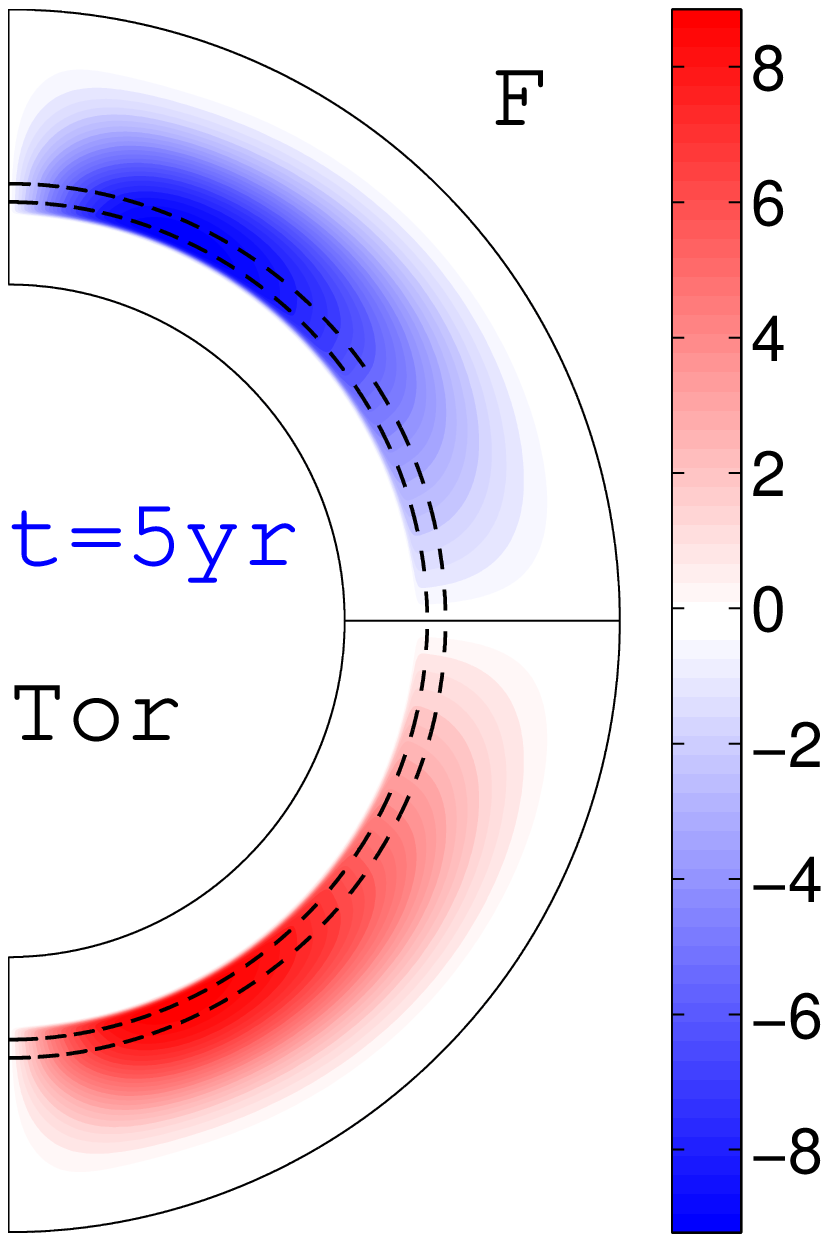}
\includegraphics[width=0.40\columnwidth]{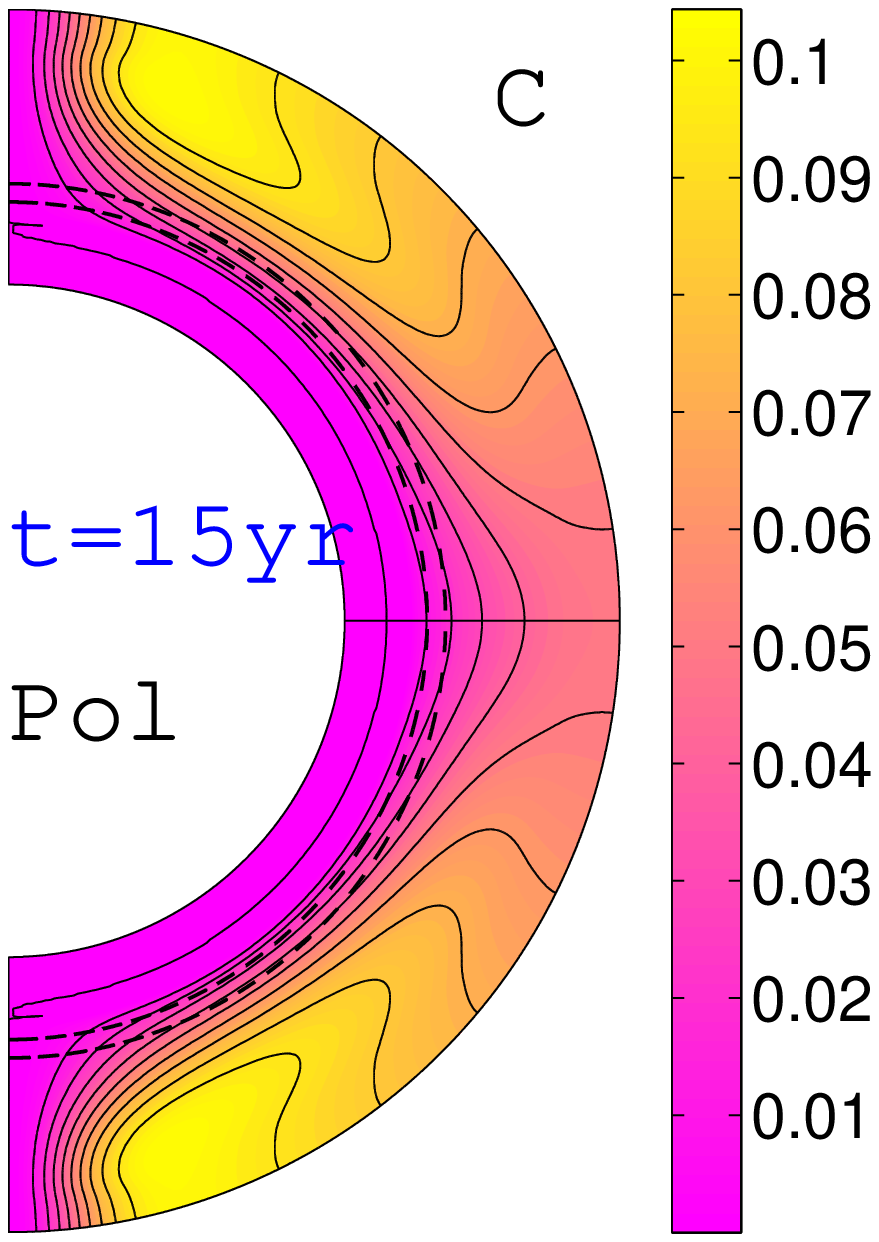}
\hspace{+0.5cm}
\includegraphics[width=0.39\columnwidth]{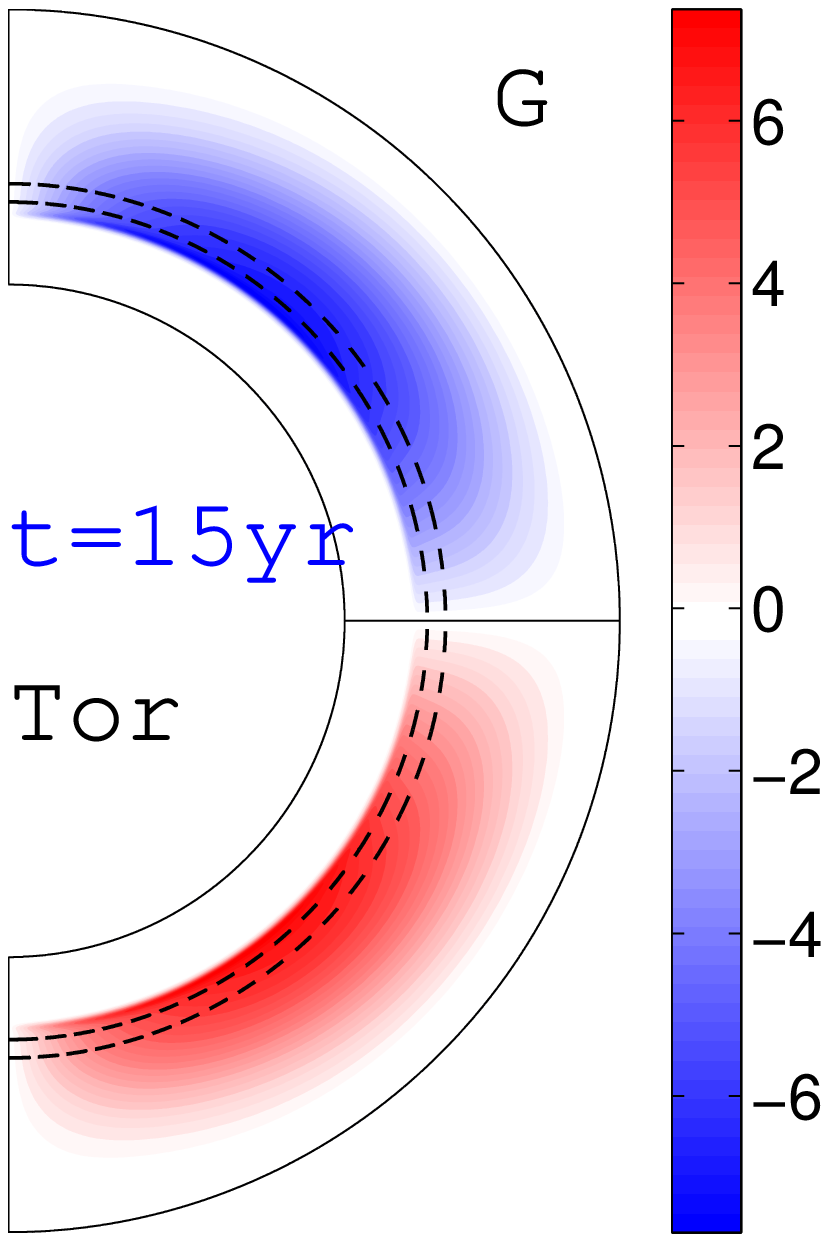}
\includegraphics[width=0.405\columnwidth]{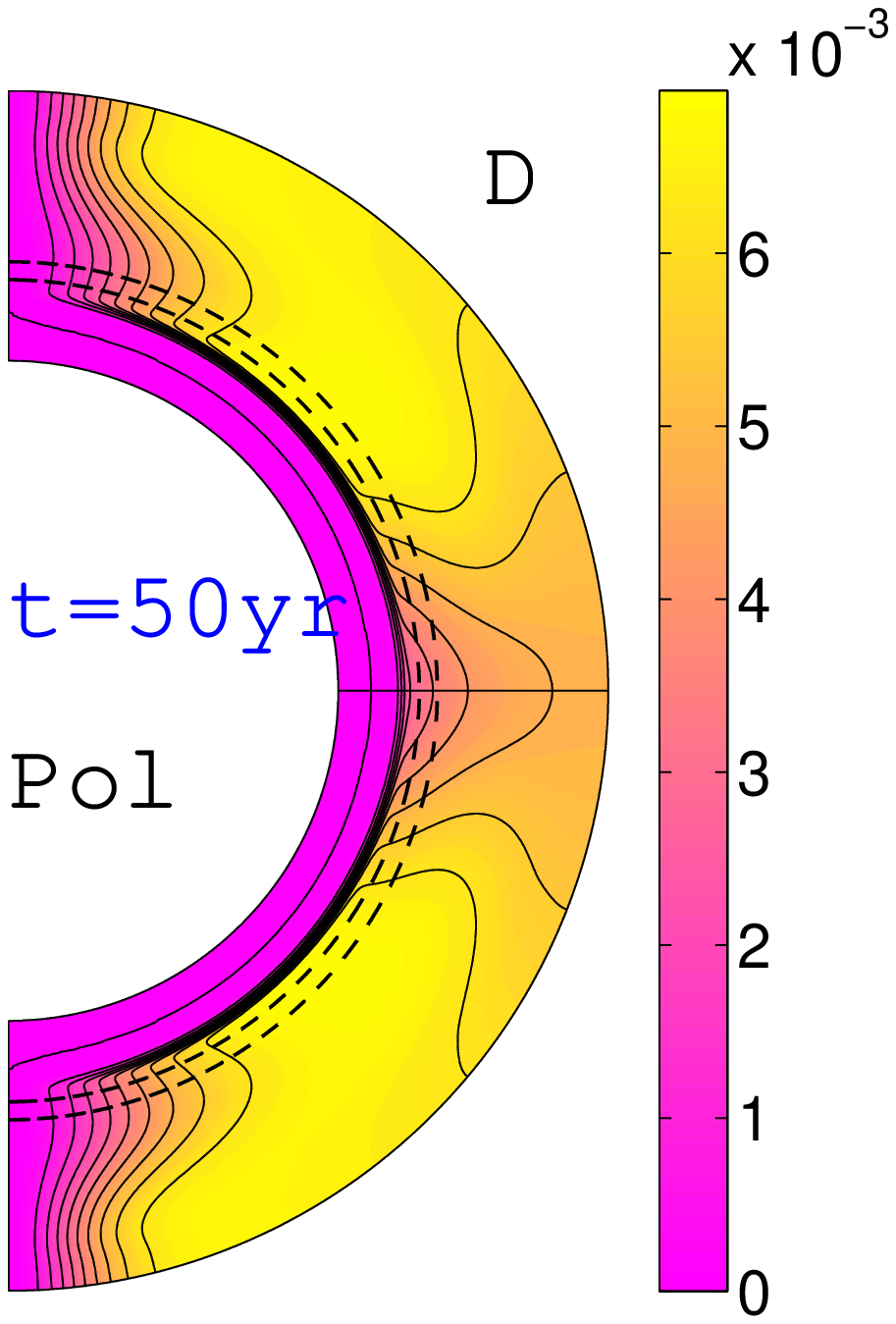}
\hspace{+0.5cm}
\includegraphics[width=0.385\columnwidth]{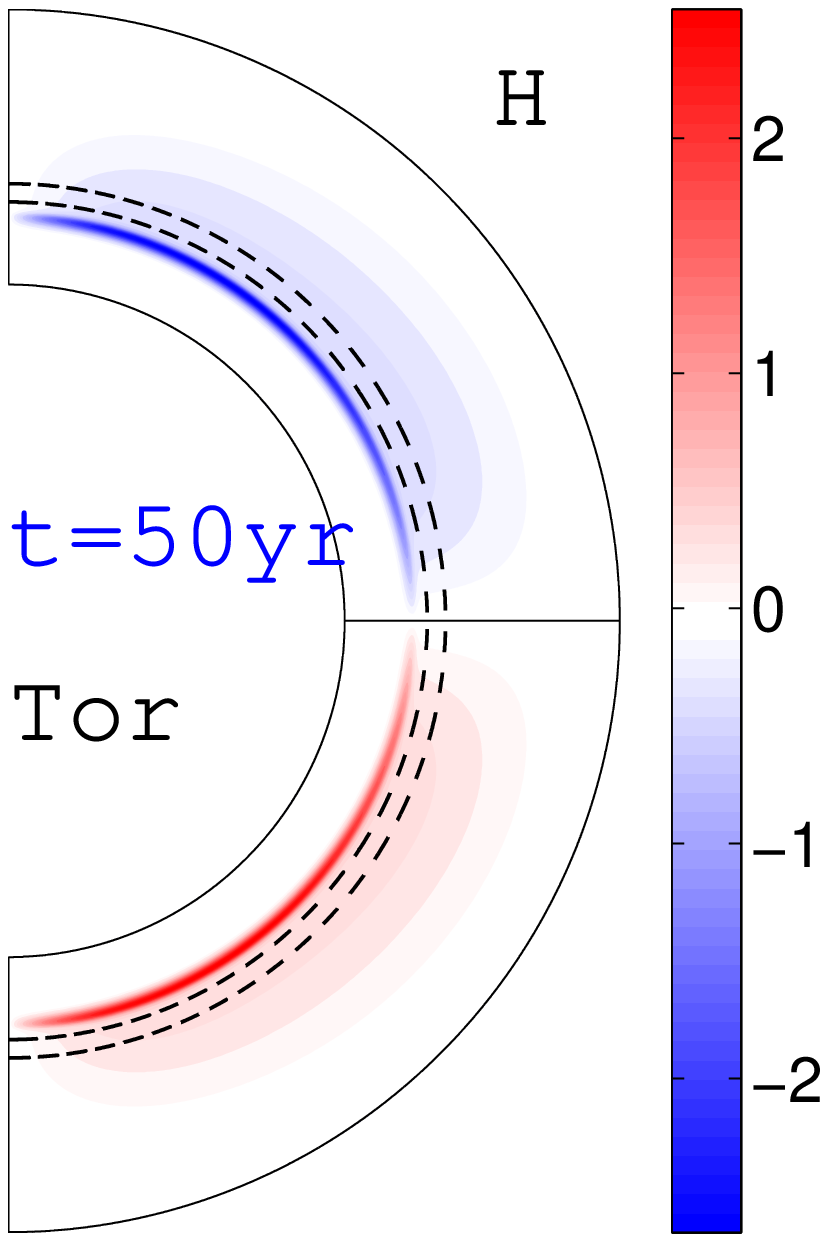}
\caption{Snapshots of poloidal field defined by the contours of constant 
$r \sin\theta A$ (left panels)
and the toroidal field (right) 
in $r$-$\theta$ planes at times $t=0$, 5, 15 and 50 years.
In this simulation
$\etab = 5 \times 10^{11}$~\cmss,
$\etas = 3 \times 10^{12}$~\cmss, \vp\ = 0~\mps\ and $\alpha_0 = 0$~\mps.
}
\label{decayp0}
\end{figure}

To get an insight into the behavior at high magnetic diffusivity, 
we begin by studying the decay of an initial poloidal 
field
without putting any magnetic pumping. The initial 
field\footnote{The conclusion of this study is independent of the choice of the 
initial poloidal field.}
is shown in panels~A of \Fig{decayp0}.
We then follow the evolution according to 
\Eqs{eqpol}{eqtor} with $\alpha_0=0$, i.e., with no source for the poloidal field in \Eq{eqpol}.
The left panels of \Fig{decayp0} show three snapshots after 5, 15 and 50~years 
from the beginning of the simulation. We observe that most of the poloidal field 
diffuses away across the surface and only in about 15~years the field has decreased by 
an order of magnitude. 

Because of the source term in the toroidal field equation, the
$\Omega$ effect, the toroidal field is generated from the poloidal field
and the snapshots after 5, 15 and 50~years are shown on the right panels 
of \Fig{decayp0}.
We observe that the latitudinal differential rotation produces a significant amount
toroidal field in the bulk of the CZ. Therefore after 5~years we have a substantial
field. However as the diffusivity is much stronger in the upper layer ($\etas=3\times10^{12}$~\cmss\ for $r\ge0.95R$),
toroidal field is much weaker there as compared to the deeper layer.
We also observe much weaker field near the equator at all radii because of
the diffusion of opposite polarity fields from opposite hemispheres.
Along with the diffusion of the toroidal field, its source, the poloidal
field itself is also decaying (compare panels A-D). Consequently
the toroidal field is not amplified beyond about 5 years, rather it starts to decrease 
in time. This can be seen in panels G and H after 15 and 50 years, respectively.

\blue{
%To make a quantitative estimates of toroidal magnetic flux loss across boundaries, 
%we begin by applying Stoke's theorem in the surface integral the induction equation,
%and obtain the rate of toroidal flux loss in the northern hemisphere \citep{Cameron15},
%$ \dot{\Phi}_{\it tor} = - \oint \! \eta {\bf \nabla} \times \bf{B} \cdot d{\it l} $.
%(Sources of magnetic flux are not considered in above).
%Here d$l$ is the line element enclosing the contour covering the northern hemisphere.
The loss of magnetic flux becomes more transparent when we monitor the net toroidal flux diffusion rates across the solar surface:
$\dot{\Phi}^{\it S}_{\it tor} = \lvert  \eta(R) \int_{0}^{\pi/2} \frac{\partial}{\partial r} (r B) d\theta  \rvert $ 
and across the equator:
$\dot{\Phi}^{\it E}_{\it tor} =  \lvert  \int_{0.55R}^{R} \eta(r)(r\sin\theta)^{-1} \frac{\partial}{\partial \theta} (\sin \theta B) dr  \rvert $,
%The rates of these diffusive fluxes (in arbitrary units), 
%along with the net toroidal flux per unit time for the northern hemisphere (in arbitrary unit) 
for the northern hemisphere (in arbitrary unit) 
which are displayed in \Fig{fig:fluxes}. We note that fluxes  
across the lower boundary and at the pole are not significant
and we do not present them here.
%; see \citep{Cameron15} for details.
}

\begin{figure}
\centering
\includegraphics[width=0.40\columnwidth]{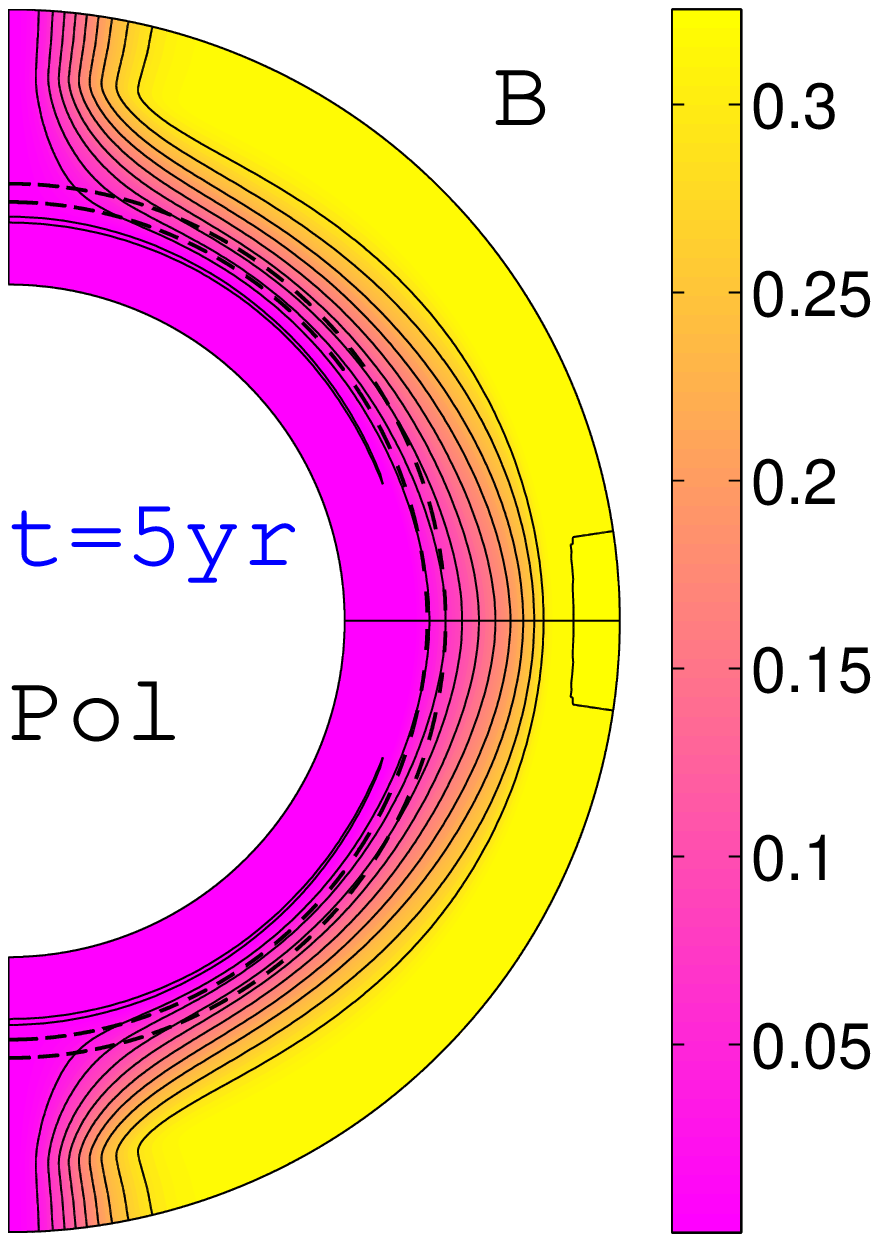}
\includegraphics[width=0.40\columnwidth]{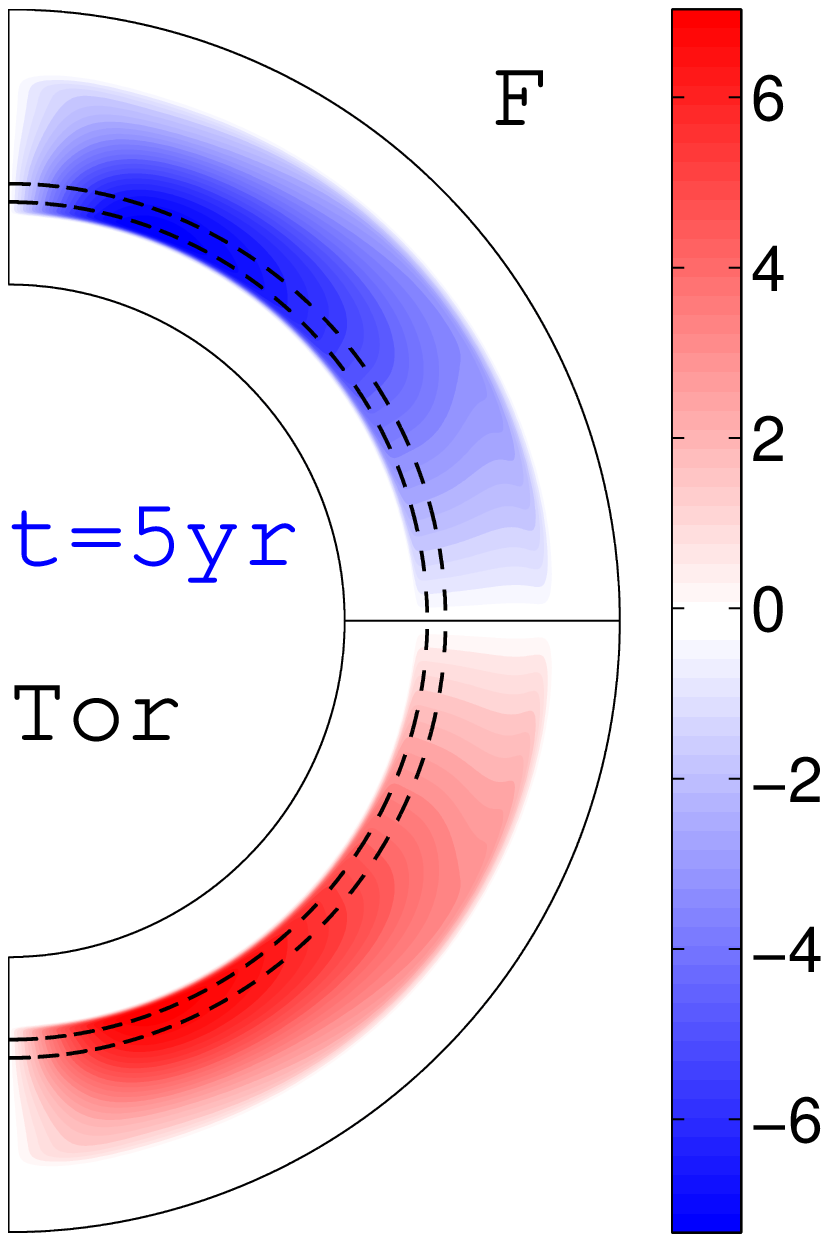}
\includegraphics[width=0.40\columnwidth]{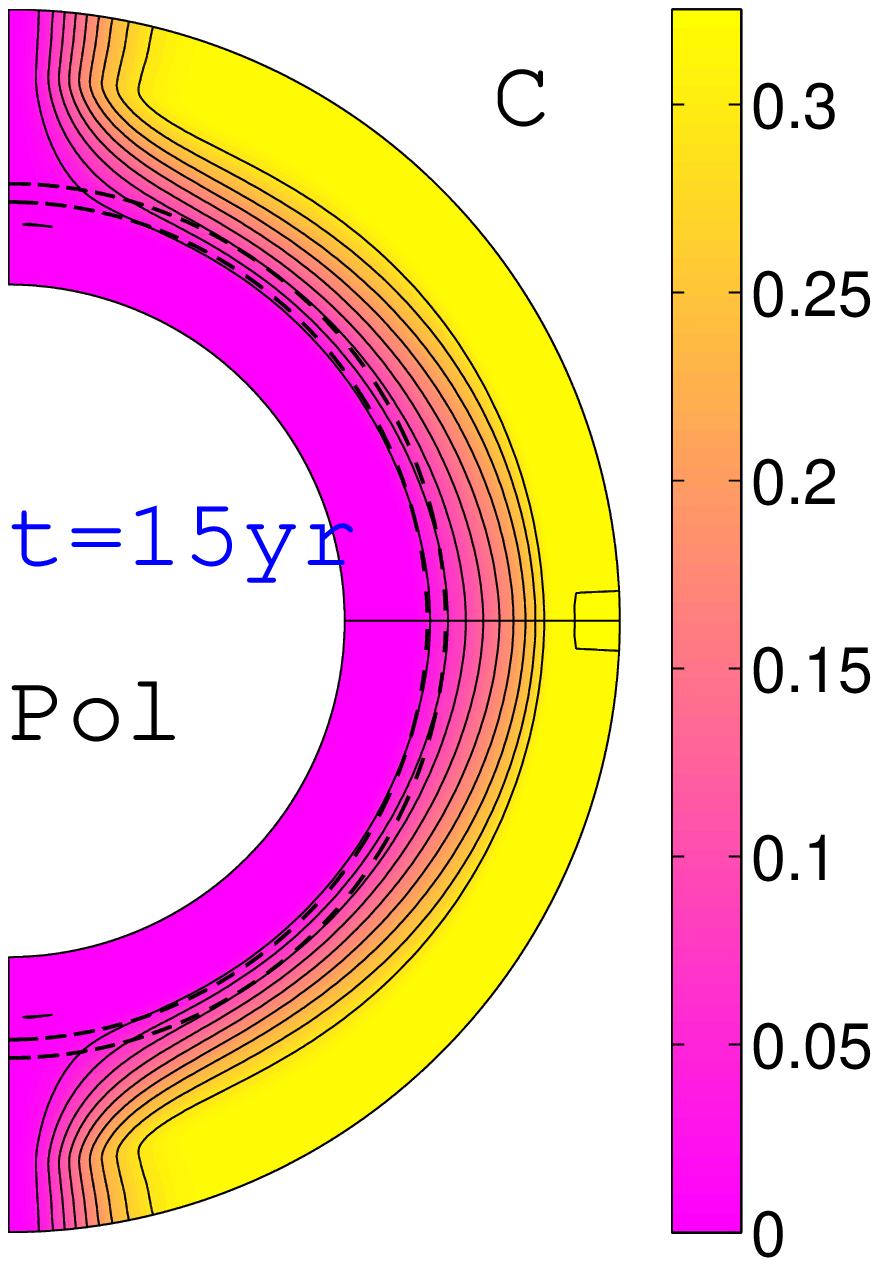}
\includegraphics[width=0.40\columnwidth]{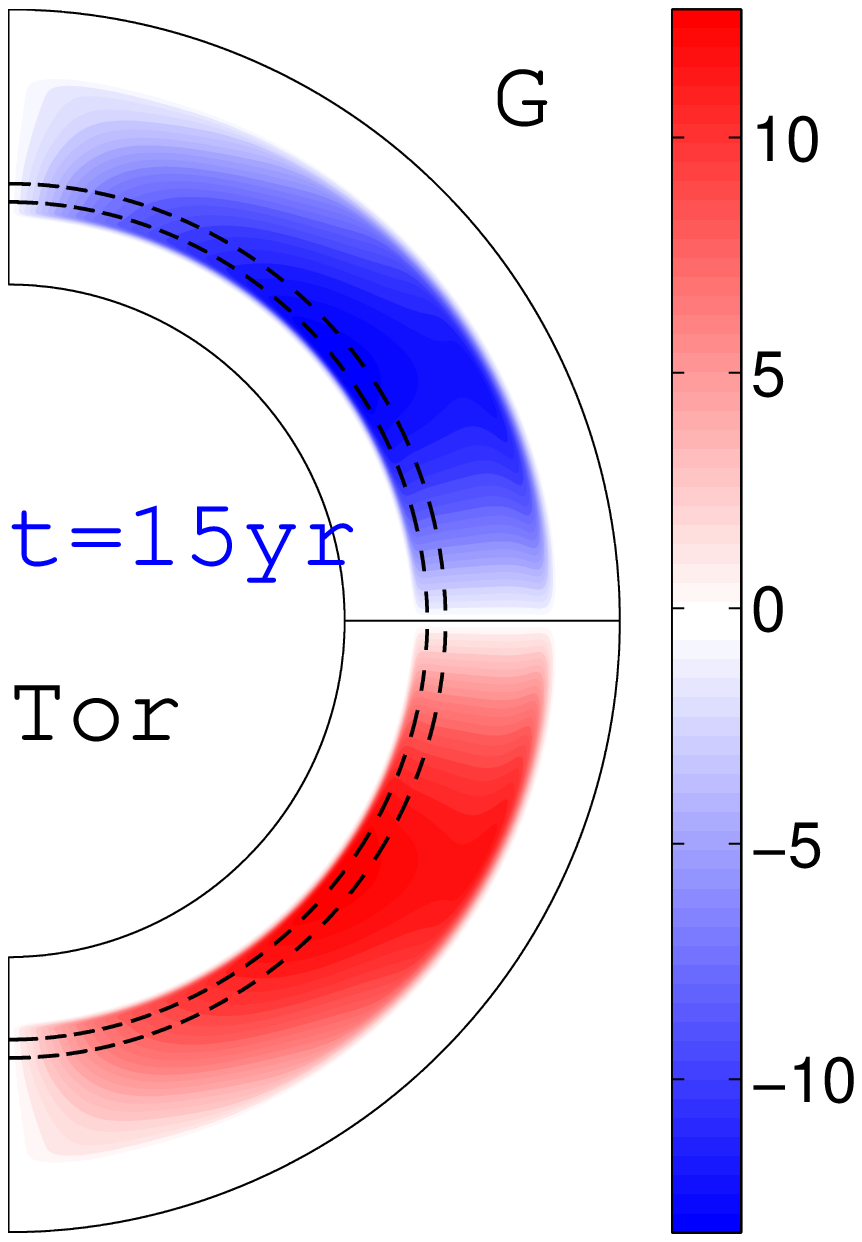}
\includegraphics[width=0.40\columnwidth]{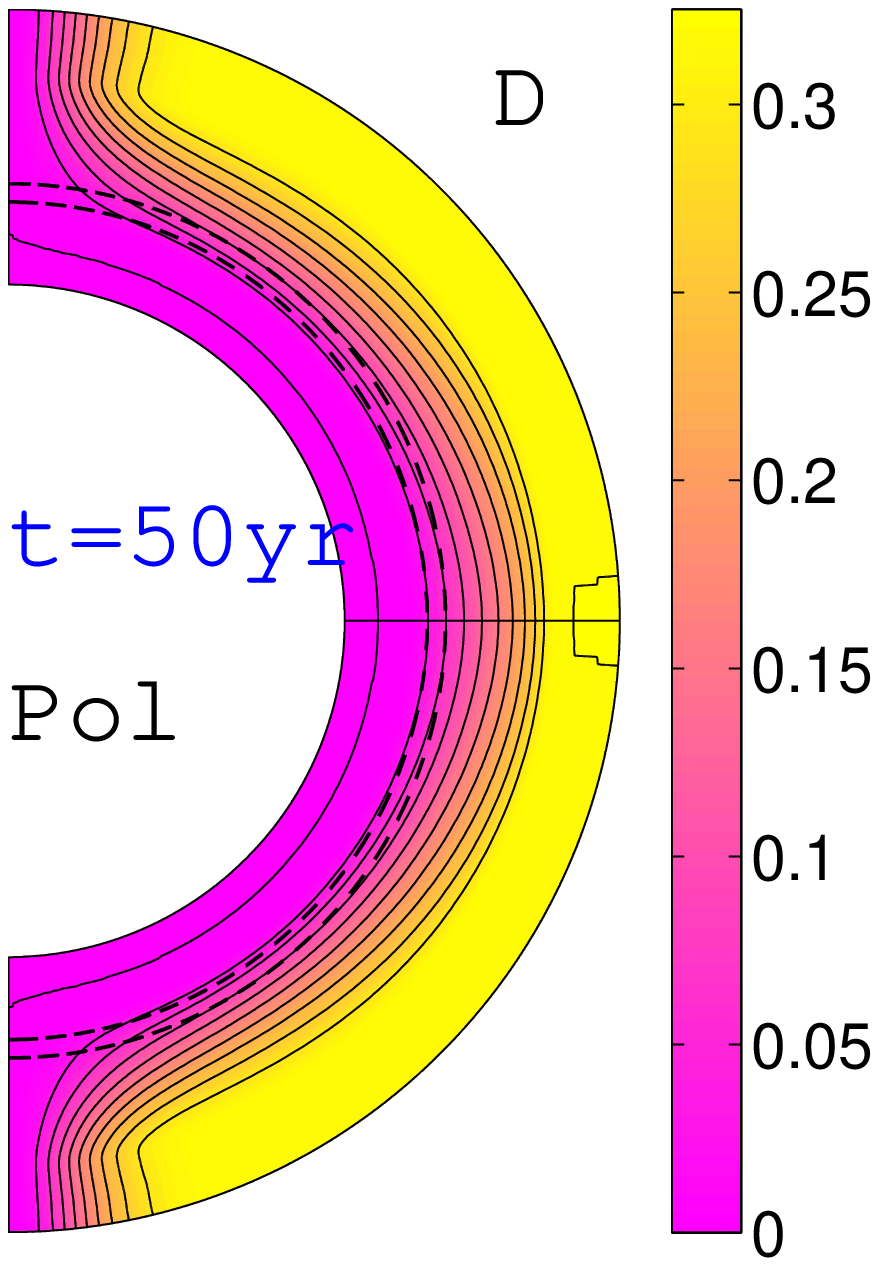}
\includegraphics[width=0.40\columnwidth]{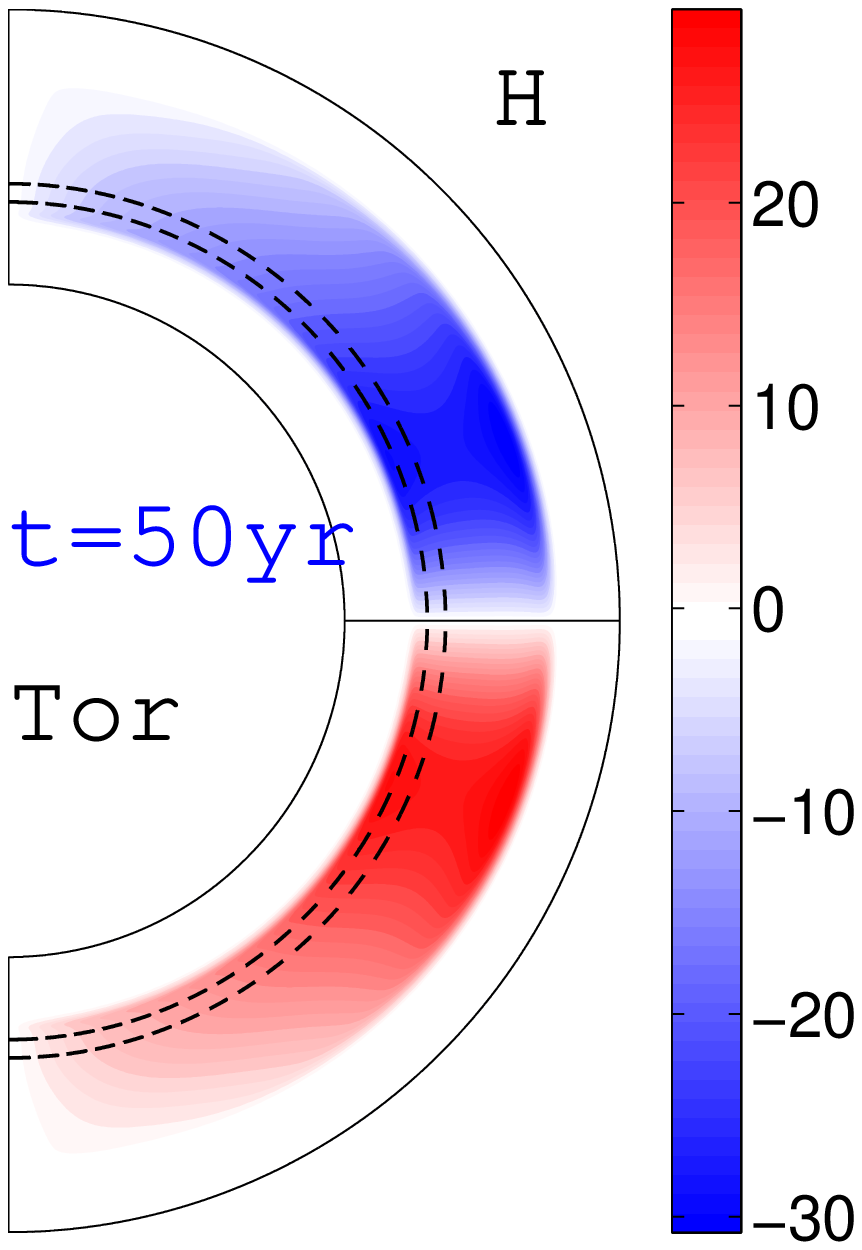}
\caption{Same as \Fig{decayp0} but at times $t=5$, 15 and 50 years 
(i.e., panels A and D are not shown here) and radial pumping \vp\ = 
35~\mps\ for $r\ge0.9R$.}
\label{decayp35}
\end{figure}

The knowledge we gain from above experiment is that the diffusion of fields
across the solar surface is extremely important for the dynamics in the high diffusivity regime.
We conjectured that a magnetic pumping in the upper layer of the CZ could inhibit 
the diffusion of the horizontal fields across the surface and radically change the dynamics.
To test this idea, we repeated the previous numerical experiment with 35~\mps\ pumping operating above $0.9 R$.
Snapshots of poloidal and toroidal fields at 5, 15 and 50~years from this simulation
are shown in \Fig{decayp35}. In comparison to the initial strength (shown in \Fig{decayp0}A), we observe that 
after 5 years the poloidal field has reduced by some extent because of the diffusion across the equator.
As expected from \citet{Ca12}, the evolution of the poloidal field is very slow after this time;
compare panels B-D in \Fig{decayp35}. The difference in poloidal fields between the simulations with and 
without pumping is due to the strongly reduced diffusion across the surface.
We further notice that the poloidal field in the near-surface shear
layer where the pumping is operating becomes 
largely 
radial. 
The poleward meridional flow
advects this field to high latitudes and once this field reaches there, 
there is no way to escape. The poloidal field remains in high latitudes
for several thousands of years, which is confirmed by continuing 
this simulation for a long time.

The toroidal field is steadily generated as before, but now its source---the poloidal 
field---is no longer decaying; see panels F-H in \Fig{decayp35}. 
However, although the toroidal field is not allowed to diffuse across the surface
because of the inward pumping, the diffusion across the equator and at poles are 
%still there.
\blue{
still there; see the black dash line in \Fig{fig:fluxes} for the rate of toroidal flux diffusion across equator.
}
Eventually, this cross-equator diffusion of toroidal flux balances the toroidal field 
generation and thereafter the system only evolves on the very long timescale ($\sim 4000$~years) 
with which the poloidal field decays.

\begin{figure}
\centering
\includegraphics[width=0.90\columnwidth]{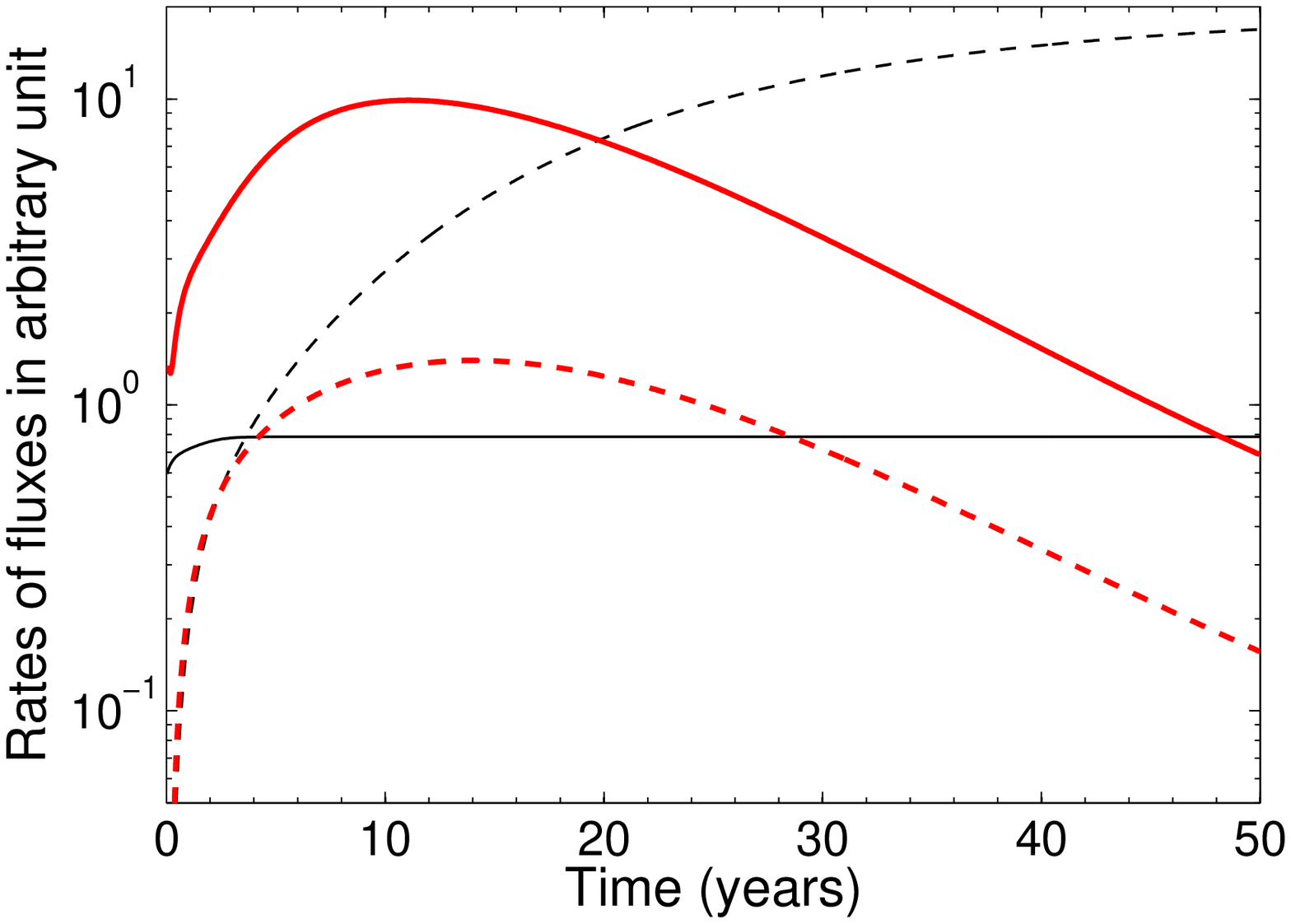}
\caption{
\blue{
Rates of net diffusive toroidal fluxes across the solar surface (solid lines),
and across the equator (dashed lines) for two simulations presented in \Figs{decayp0}{decayp35}, without 
pumping (red/thick) and with 35 m/s pumping (black/thin), respectively.
}
}
\label{fig:fluxes}
\end{figure}

\begin{figure}
\centering
\includegraphics[width=1.12\columnwidth]{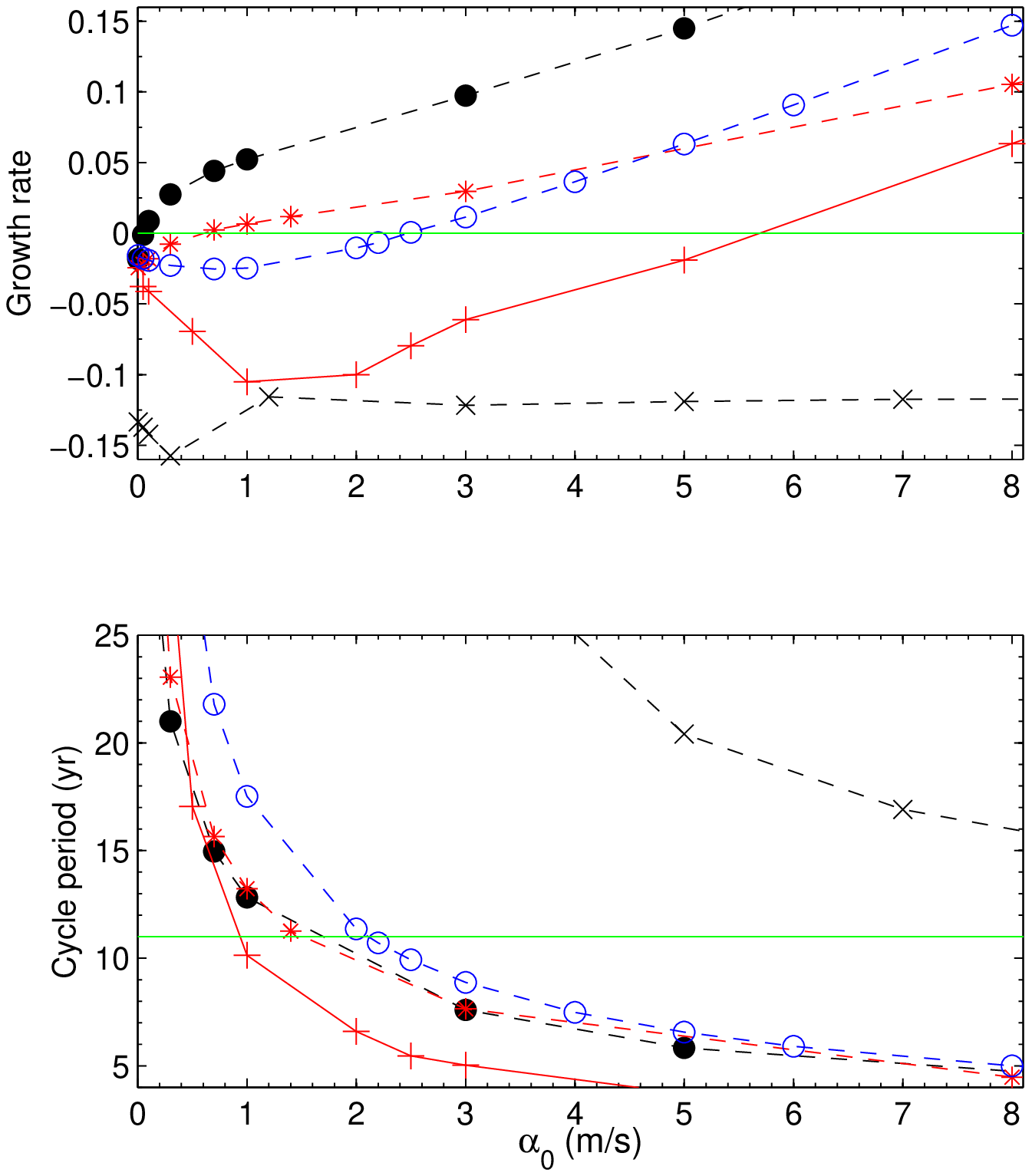}
\caption{Results using the analytic rotation profile as shown in \Fig{omgan}: 
growth rates (per year) of the net toroidal flux in one hemisphere (upper panel) and
the cycle period (lower panel) as functions of $\alpha_0$.
Crosses represent simulations with $\etab = 5\times10^{11}$~\cmss\ and 
no magnetic pumping (\vp\ = 0), while for all other symbols 
\vp\ = 35~\mps.
Filled points: $\etab = 5\times10^{11}$~\cmss, red asterisks: $\etab = 1\times10^{12}$~\cmss,
open circles: same as filled points but with no {\it equatorward} component of meridional flow,
and plus symbols: same as filled points but no \mc. 
}
\label{fig:grOmgan}
\end{figure}

\subsection{Dynamo solutions with $\alpha^{\rm BL} \neq 0$}
Now we make $\alpha_0 \neq 0$ and perform a few sets of simulations by varying 
it within [0.005--10]~\mps.
From now onward we consider the initial condition of our simulations
as $B=\sin(2\theta) \sin[\pi(r-0.7R)/(R-0.7R)]$ 
for $r > 0.7R$ and $0$ for $r\le0.7R$, while $A=0$ for all $r$ and $\theta$.
We run the simulation for several cycles so that our results are not much dependent on the initial condition.

When there is no magnetic pumping, the initial seed field decays like a damped oscillator for all values of $\alpha_0$ in this range 
(expected based on the knowledge gained in previous decay experiments). 
Crosses in \Fig{fig:grOmgan} show the field decay rates (top) and
the cycle periods (bottom). 
We notice that the cycle period decreases with increasing $\alpha_0$ 
%as expected in an $\alpha \Omega$ dynamo model. 
as expected in a 
\blue{
linear
}
 $\alpha \Omega$ dynamo model. 
However, the dependence of period with $\alpha_0$ in the FTD model 
is only mild \citep[see Eq.\ (12) of][]{DC99} possibly because of a strong B-dependent
quenching introduced in the $\alpha$ effect.

Astonishingly, when the downward pumping of sufficient strength is present, 
results change completely.
For $\alpha_0$ below a certain value, the initial field still decays due to the diffusion, however above some critical value,
we get growing oscillations. We note that in these simulations we have not considered any nonlinearity in Babcock-Leighton process to stabilize the dynamo growth.
The filled circles in \Fig{fig:grOmgan} represent the results from simulations with
35~\mps\ downward pumping. Here we notice that the periods are much shorter compared to the
simulations without pumping (as seen in the lower panel of the same figure).
There are multiple reasons for producing such a short period. 
One is the longer lifetime (slower decay) of both the poloidal and toroidal
fields due to the reduction in diffusion of flux through the photosphere as discussed above.
Another is that the pumping helps to transport the field
from the surface to the deeper layers of the CZ where the toroidal field can be 
amplified further and erupt to
produce flux emergence \citep{KN12}.

\begin{figure}
\centering
\includegraphics[width=0.90\columnwidth]{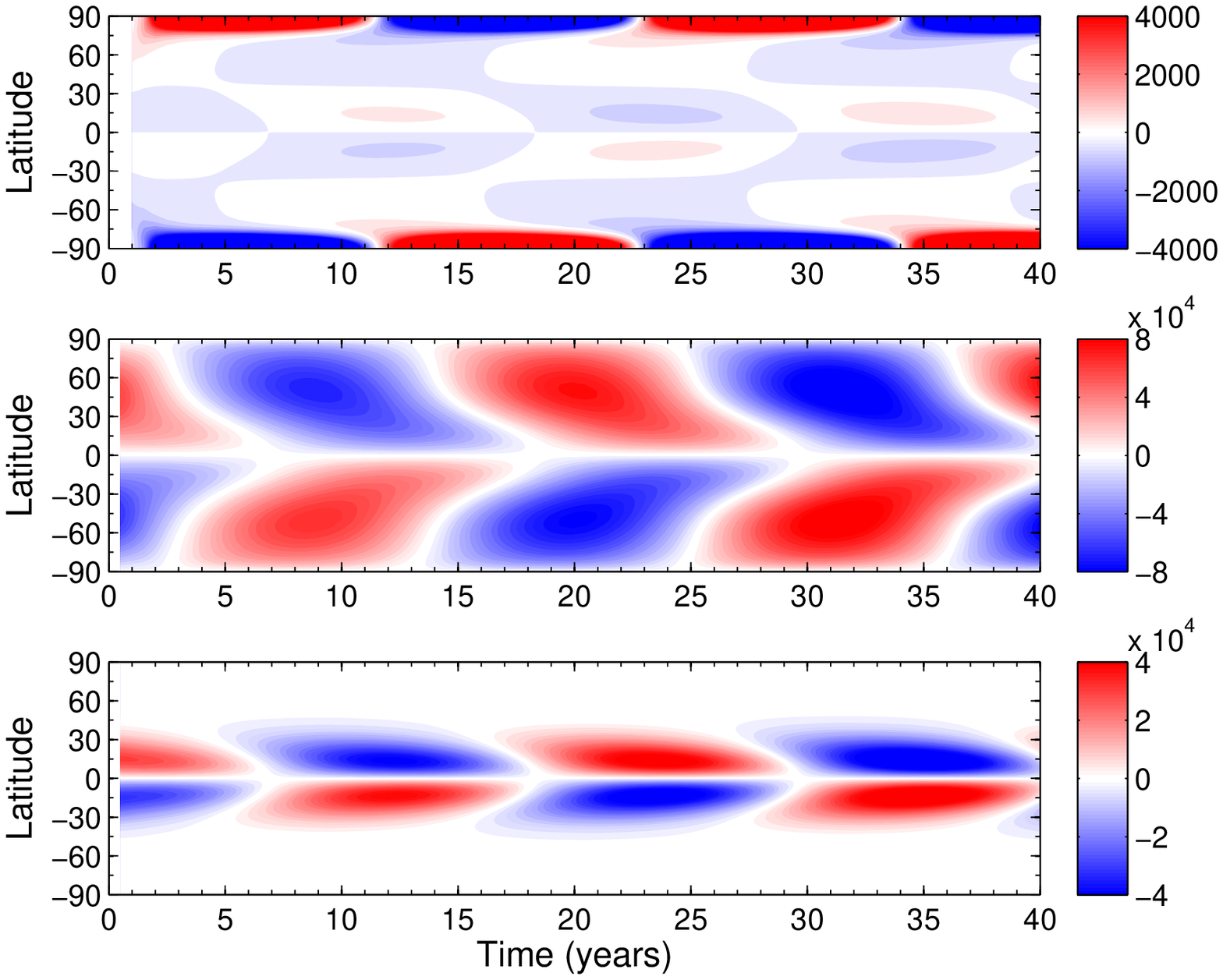}
\caption{Time-latitude diagrams of the radial field (G) on the solar surface (top), 
mean toroidal field (G) over the whole CZ (middle) and the \bl\ source term, 
\Eq{BLsource} (bottom).
In this simulation,
$\alpha_0 = 1.4$~\mps, 
$\etas = 3 \times 10^{12}$~\cmss, $\etab = 1 \times 10^{12}$~\cmss
and \vp\ = 35~\mps.
}
\label{butOman}
\end{figure}

For the parameters considered here, i.e., $\etab = 5 \times 10^{11}$~\cmss,
$\etas = 3 \times 10^{12}$~\cmss\ and \vp\ = 35~\mps,
the dynamo period is nearly 63 years at near the critical value of $\alpha_0$ for dynamo action (the growth rate for these parameters is 0.008 per year).
%Therefore, we considered a case with a diffusivity $\etab$ of $1\times10^{12}$~\cmss,
%and again perform a set of simulations at different values of $\alpha_0$.
Therefore, we perform a set of simulations at a higher value of diffusivity $\etab=1\times10^{12}$~\cmss.
%and again perform a set of simulations at different values of $\alpha_0$.
Red asterisks in \Fig{fig:grOmgan} represent these.
%From these simulations, we get a nearly stable solution (growth rate = 0.01 per year) 
Here we get a nearly stable solution (growth rate = 0.01 per year) 
and 11-year period at $\alpha_0 = 1.4$~\mps.
\Fig{butOman} shows the butterfly diagrams of the radial field, 
the mean toroidal field over the CZ and the \bl\ source
term from this simulation. This figure displays most of the regular 
features of solar magnetic cycle
e.g., the regular polarity reversal, correct phase relation between the radial and toroidal
fields, the equatorward migration of toroidal field at low
latitudes and the poleward migration of the surface radial field.

\begin{figure}
\centering
\includegraphics[width=1.0\columnwidth]{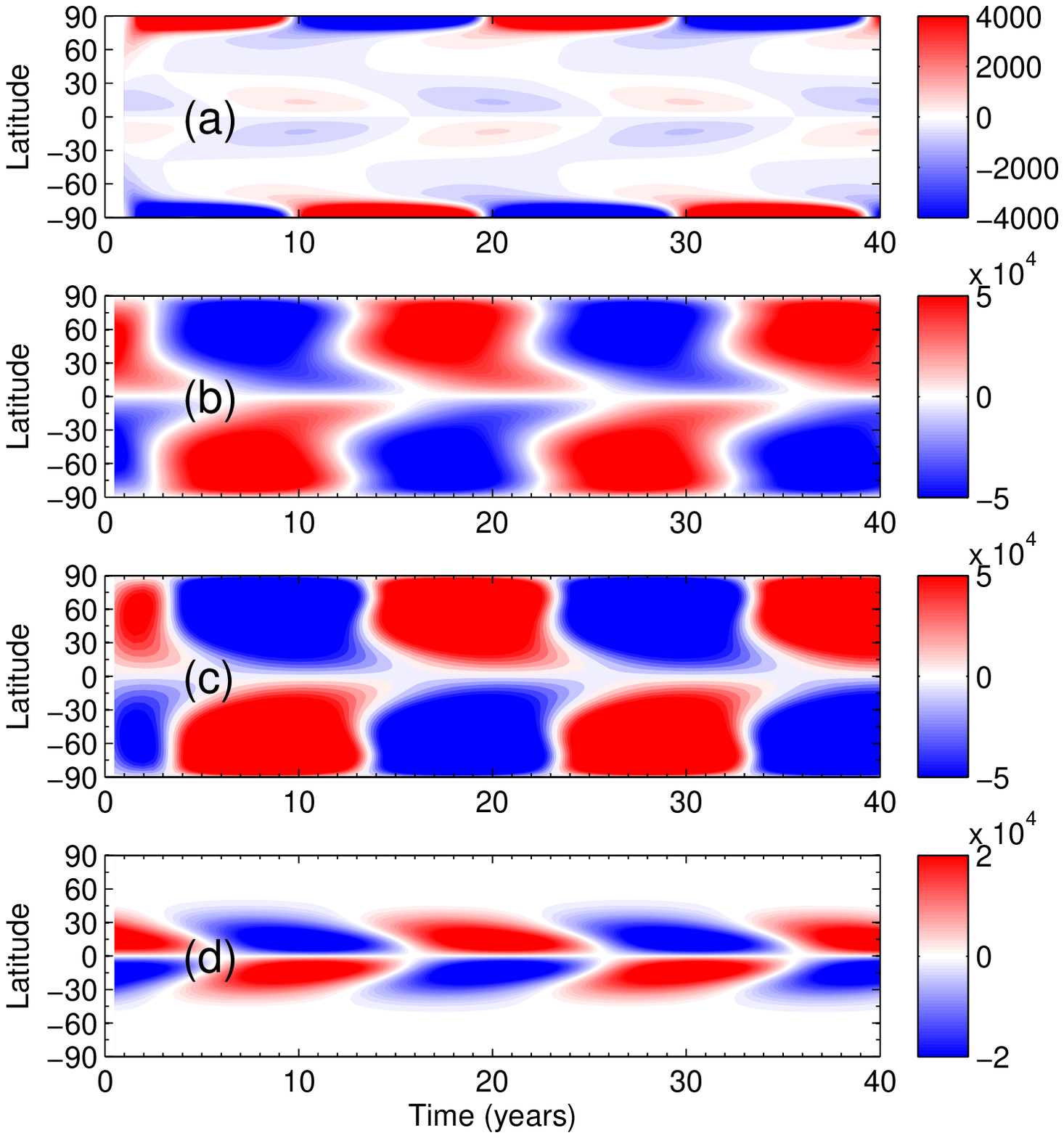}
\caption{
Same as \Fig{butOman} except an additional panel (c),
showing the toroidal field over tachocline
and with different model parameters:
$\alpha_0 = 2.5$~\mps, $\etab = 5 \times 10^{11}$~\cmss\ and
{\it no equatorward component} in the meridional flow.
(For better visibility of the weak fields, colorbars
are clipped at displayed ranges.)
}
\label{butOmannoeqMC}
\end{figure}
\begin{figure}
\centering
\includegraphics[width=1.0\columnwidth]{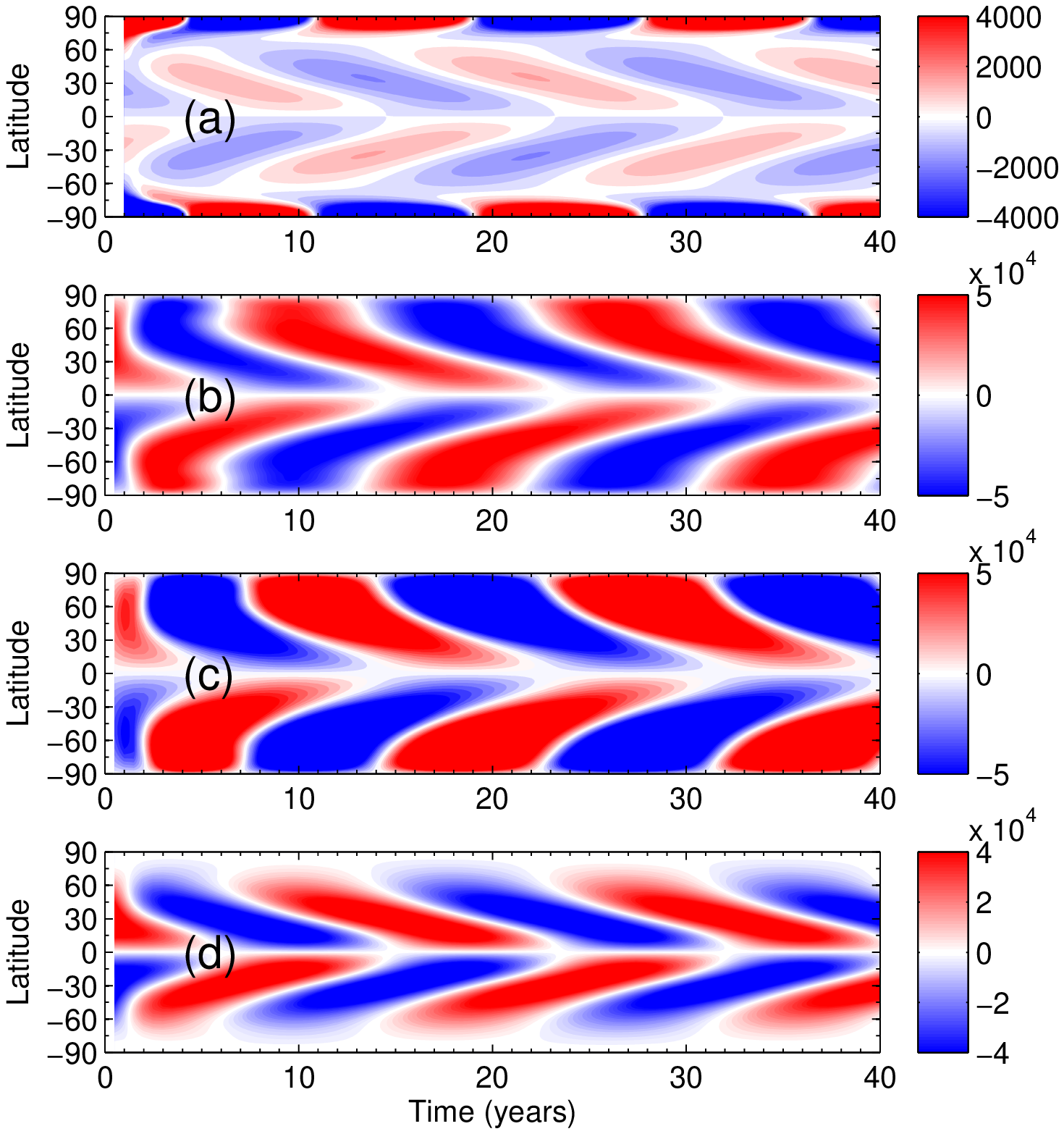}
\caption{Same as \Fig{butOmannoeqMC} but in this simulation 
$f_\alpha = \cos\theta \sin^{2}\theta$ (i.e., $n=2$ in \Eq{eqlal}) is used.
}
\label{butOmannoeqMCcs2}
\end{figure}

To understand the importance of the equatorward \mf\ we repeat the set of simulations at $\etab = 5\times10^{11}$~\cmss\ and \vp\ = 35~\mps\
by making the equatorward \mf\ (in the lower CZ) to zero. 
We emphasize that this setup is designed to disentangle the importance of equatorward \mf\ and the shear in the near-surface layer.
However, we retain the poleward component of \mf\ in the upper CZ,
which is crucial in producing the observed poleward migration of the radial field.
To avoid discontinuity, we do not change the radial component of the meridional flow
and it remains as it is in the full meridional cell. This component
is anyway very weak to affect the result.
Instead of making this unphysical setup, we could, of course, consider a shallow meridional flow
here.
Nevertheless, this would contribute
an equatorward flow at some depth and help to produce equatorward migration in toroidal field.
Hence, to exclude the effect of the equatorward flow and to observe the consequence of the dynamo wave alone,
we artificially cease the equatorward return flow.
It is formally consistent with the kinematic mean-field equations 
where the mean meridional flow is prescribed. We comment that
usually the meridional flow is assumed to satisfy $\nabla \cdot (\rho{\bf{v_p}})=0$ 
on the basis that the flow is steady and conserves mass, 
however in the mean-field framework mass conservation of a steady mean flow yields,
$\nabla \cdot (\rho{\bf{v_p}})= - \nabla \cdot (\overline{\rho'{\bf{v_p'}}})$, 
where primes indicate the fluctuating components and the
overbar represents the resulting mean. The usual habit of using $\nabla \cdot (\rho{\bf{v_p}})=0$ reflects the fact that
$- \nabla \cdot (\overline{\rho'{\bf{v_p'}}})$ is unprescribed by the model and treating it as anything other than zero introduces more
parameters into a model that already has a large number of free parameters. For our purposes, 
setting the return flow to zero is not inconsistent in terms of the model. 

The open circles in \Fig{fig:grOmgan} represent these simulations. 
Interestingly, we again find growing solutions for $\alpha_0 > 2.45$~\mps.
In \Fig{butOmannoeqMC}, we show the butterfly diagrams for $\alpha_0 = 2.5$~\mps\ 
(for which the growth rate = 0.0007 per year).
The striking result is that we get a clear equatorward migration at low latitudes. As there is no equatorward \mf\ in these simulations, this 
equatorward migration is caused by the negative shear exists in the near-surface layer and the positive $\alpha^{\rm BL}(r,\theta)$.
This is a dynamo wave obeying the Parker--Yoshimura sign rule \citep{Pa55,Yo75}. 
This result is in agreement with \citet{GD08} (see their Sec.~4) who 
found equatorward dynamo wave near the surface by including a NSSL.
However, they need much weaker pumping ($< 1$~m~s$^{-1}$) than used here
because the near-surface diffusivity is much smaller in \citet{GD08} than we use.

As our $\alpha^{\rm BL}(r,\theta)$ is nonzero below $\approx \pm 30^{\circ}$ latitudes (where we see sunspots), 
we observe an equatorward migration only below $\approx \pm 30^\circ$ latitudes.
Therefore when we extend the latitudinal width of $\alpha^{\rm BL}(r,\theta)$ 
by changing $f_\alpha$ from $\cos\theta \sin^{12}\theta$ to
$\cos\theta \sin^{2}\theta$, we observe much prominent equatorward migration 
starting from higher latitudes 
than we found earlier; see \Fig{butOmannoeqMCcs2}.
This equatorward migration caused by the dynamo wave is not
localized near the surface, rather it is propagated throughout the CZ; see \Fig{butOmannoeqMC}(c) and \Fig{butOmannoeqMCcs2}(c).
This is because the CZ is strongly radially coupled due to the high diffusivity
\blue{
as well as due to strong pumping operating in the upper part of the CZ.
}

Another feature to note is that the surface radial field 
(top panels of \Figs{butOmannoeqMC}{butOmannoeqMCcs2}) 
also propagates equatorward---as a consequence of dynamo wave---in 
low latitudes where $\alpha^{\rm BL}(r,\theta)$ is nonzero, 
but shows a weak poleward migration in high latitudes---as 
a consequence of poleward meridional flow.
Therefore, if the equatorward migration of butterfly wings of the Sun is really caused
by the dynamo wave, then the $\alpha$ effect must be concentrated in low latitudes only
(as we choose in our study by setting $f_\alpha = \cos\theta \sin^{12}\theta$), 
otherwise the poloidal field as a consequence of dynamo wave 
will migrate equatorward even in high latitudes 
which is not observed.

We further note that in these simulations, although the equatorward 
return flow was artificially switched off, its 
poleward component in the upper CZ was retained. 
This poleward meridional flow 
is crucial to transport the poloidal field generated through the 
\bl\ process at low latitudes. Therefore, if we switch off this 
surface-poleward flow, 
then the cycle period becomes shorter and the dynamo decays at these parameters.
For details, we consider red plus symbols in \Fig{fig:grOmgan} 
which represent the set of simulations at
$\etas = 5 \times 10^{11}$~\cmss\ and \vp\ = 35~\mps\ but no meridional flow.
We observe that dynamo is growing only when $\alpha_0 \ge$ $6$~\mps\ but
cycle periods are shorter. 
Thus in dynamo wave solution, although meridional flow is
not required to produce the equatorward migration of Sun's butterfly wings, 
the poleward surface flow is crucial to get a correct cycle period in the Sun.

%===============================================================================
\section{Dynamo solutions with observed rotation profile}
\label{sec:Omobs}

We now perform simulations with the helioseismic data for angular velocity 
as shown in \Fig{omgob}(a). 
For the poloidal source, the $\alpha$ effect, we either relate it 
to the mean toroidal field in the CZ or in the tachocline.

\begin{figure}
\centering
\includegraphics[width=1.09\columnwidth]{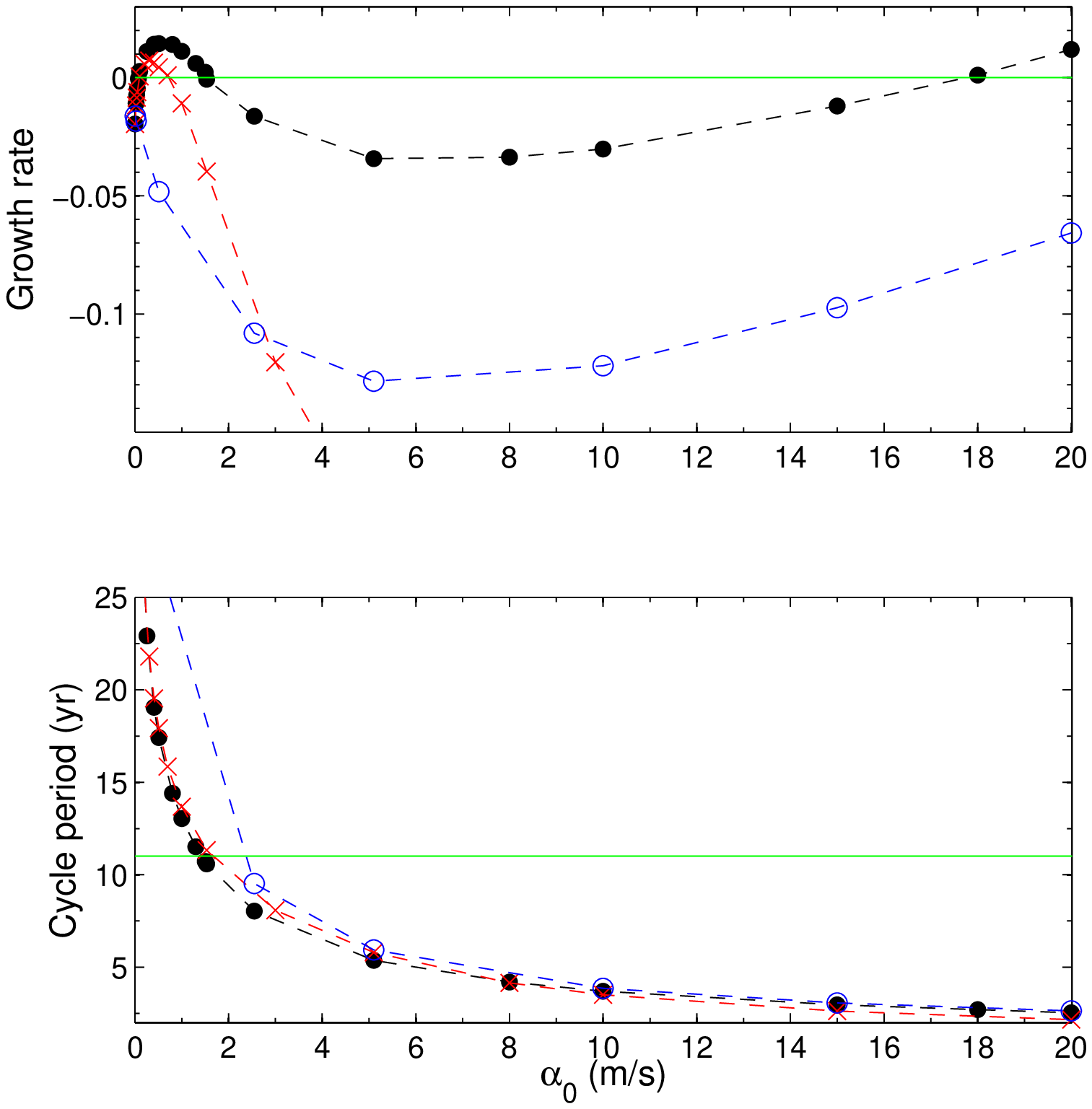}
\caption{Results using the {\it observed} differential rotation as shown in \Fig{omgob}: 
Growth rates of toroidal flux (top) and the cycle periods (bottom) as functions of $\alpha_0$
from simulations with \vp\ = 35~\mps\ 
and $\etab = 5\times10^{11}$~\cmss\ (filled points). 
Red crosses represent simulations in which the radial shear in the near-surface shear 
is artificially set to zero, while open circles represent simulations with 
no equatorward meridional flow; 
everything else remain same in these two sets.
}
\label{fig:grOmobs}
\end{figure}

\begin{figure}
\centering
\includegraphics[width=0.90\columnwidth]{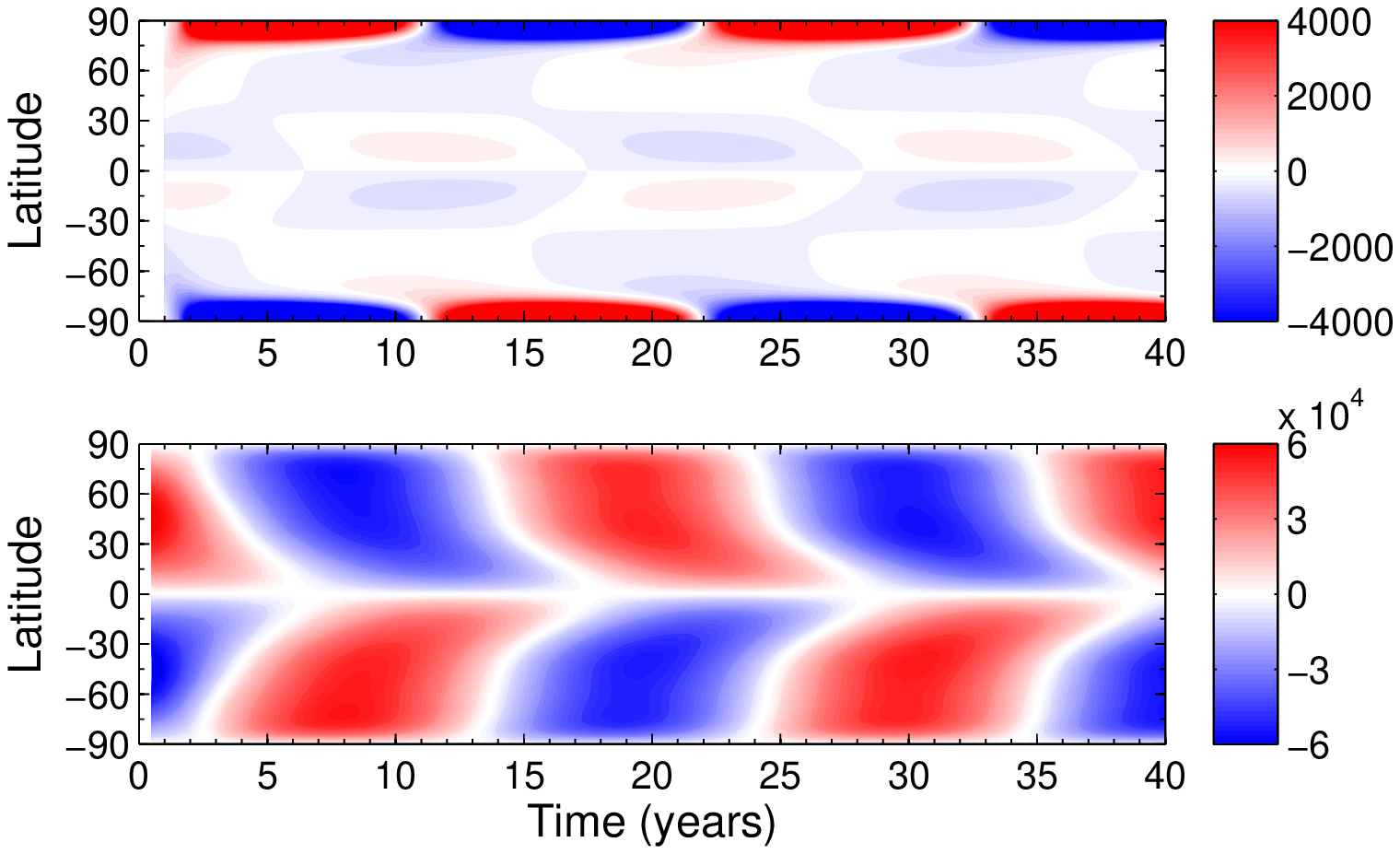}
\caption{
Results using {\it the observed} differential rotation, $\alpha_0 = 1.5$~\mps, \vp\ = 35~\mps\
and $\etab = 5\times10^{11}$~\cmss.
Time-latitude diagrams of the radial field on the solar surface (top) 
and the mean toroidal field over the whole CZ (bottom).
}
\label{butOmobs}
\end{figure}

%----------------------------------------------------------------------------
\subsection{Poloidal source relating to the mean toroidal field in CZ}
Similar to previous studies we perform a few sets of simulations with  $\etab = 5\times10^{11}$~\cmss\ and \vp\ = 35~\mps\ by varying $\alpha_0$
within $[0.003-20]$~\mps.  The filled points in \Fig{fig:grOmobs} show the corresponding
growth rates and dynamo periods. Here we see a non-monotonous behavior for growth rate.
For $0.1 \le \alpha_0 \le 1.5$~\mps\ we get positive growth, while for the intermediate value,  
$1.53 \le \alpha_0 \le 18$~\mps, we have a region of unexpected negative growth. In \Fig{butOmobs}, we show the solution for $\alpha_0 = 1.5$~\mps.
Again this figure reproduces the basic features of the solar magnetic cycle.

Now we consider the other values of $\alpha_0$ (filled circles) in \Fig{fig:grOmobs}.
With the increase of $\alpha_0$, the production of poloidal field is faster which makes the field reversal quicker,
thereby making the cycle period shorter (as expected).

The decay for intermediate values of $\alpha_0$ can be understood in the following way. In our model, the toroidal field is produced in
the bulk of the CZ with the strongest production being at the high latitudes, while the poloidal field
is produced from the toroidal field in low latitudes below $30^\circ$ (where sunspot eruptions occur).
At lower values of $\alpha_0$, the cycle period is long enough for the toroidal flux to be transported to the low latitudes to
feed the flux emergence and subsequently the poloidal field generation. This transport of the toroidal flux is done by the equatorward flow 
because the advection time, say from latitude $60^\circ$ to $15^\circ$ is about 10~years for a 1~\mps\ meridional flow, while the diffusion time is about 100~years for
$\etab = 5\times10^{11}$~\cmss. 
However at higher $\alpha_0$, when the cycle period becomes shorter than
this advection time, the advection of toroidal field becomes less important.
Hence the toroidal field at low latitudes
decreases to make the production of the poloidal field weaker and consequently
decay the oscillation.
However, when $\alpha_0$ is increased to a very large value, the cycle
period does not decrease much (see \Fig{fig:grOmobs} bottom), but a stronger
$\alpha_0$ can produce a sufficient poloidal field 
even at a weaker toroidal field
to make the dynamo growing again.

When we increase the diffusivity $\etab$ above $7.5\times10^{11}$~\cmss, we 
get decaying solutions for all values of $\alpha_0$ tested.
We note that in the previous section, where we used the simplified profile for $\Omega$,
we had a growing dynamo solution with 11-year
period even with a diffusivity of $10^{12}$~\cmss.
This provides a measure of the
sensitivity of dynamo action to the details of the differential rotation, pumping and diffusivity. We can definitely state that
dynamo action is possible with turbulent magnetic diffusivities $\etab$ up to about $7.5\times10^{11}$~\cmss, however there are
enough free parameters and choices in the model to prevent us ruling out plausible solutions with higher diffusivities.

By repeating the set of simulations with $\etab = 5\times10^{11}$~\cmss\ and \vp\ = 35~\mps\
with artificially zeroing the equatorward \mf\ near the bottom of CZ, we obtain
decaying oscillations for all values of $\alpha_0$ (open circles in \Fig{fig:grOmobs}).
The reason for this is already explained above.
When there is no flow near the bottom of the CZ,
the toroidal field from high to low latitudes can only be reached
by the diffusion (with diffusion time about 100 years).
Therefore with the decrease of cycle period due to the increase of $\alpha_0$,
the diffusion of the field from high to low latitudes decreases
which effectively makes the
production of the poloidal field weaker and causes the dynamo to decay.
However it can again produce growing solutions at larger values of $\alpha_0$.

Finally, instead of putting the equatorward \mf\ to zero,
we now repeat the above set of simulations by putting 
the radial shear in the near-surface layer to zero. 
The red crosses in \Fig{fig:grOmobs} represent this set.
We obtain growing solutions for some smaller values of
$\alpha_0$ and then decaying at larger values.
This behavior is very similar to the previous set of simulations with near-surface shear 
(filled points) except in this case the dynamo growth is smaller.
Therefore the solutions with the observed differential
rotation profile at smaller values of $\alpha_0$ are FTD type, modified by the NSSL. 
At the highest values of $\alpha_0$ we have considered, the dynamo-wave solution becomes more relevant, (although the solution without
the equatorward return flow is still subcritical at this $\alpha_0$). 
%The solution with the observed rotation has a positive growth.
%We identify this solution as a dynamo-wave solution modified by the meridional return flow. 
%We note that this branch has a period much shorter
%than the solar cycle.

%------------------------------------------------------------------------------------------
\subsection{Poloidal source relating to the tachocline toroidal field}
\begin{figure}
\centering
\includegraphics[width=1.05\columnwidth]{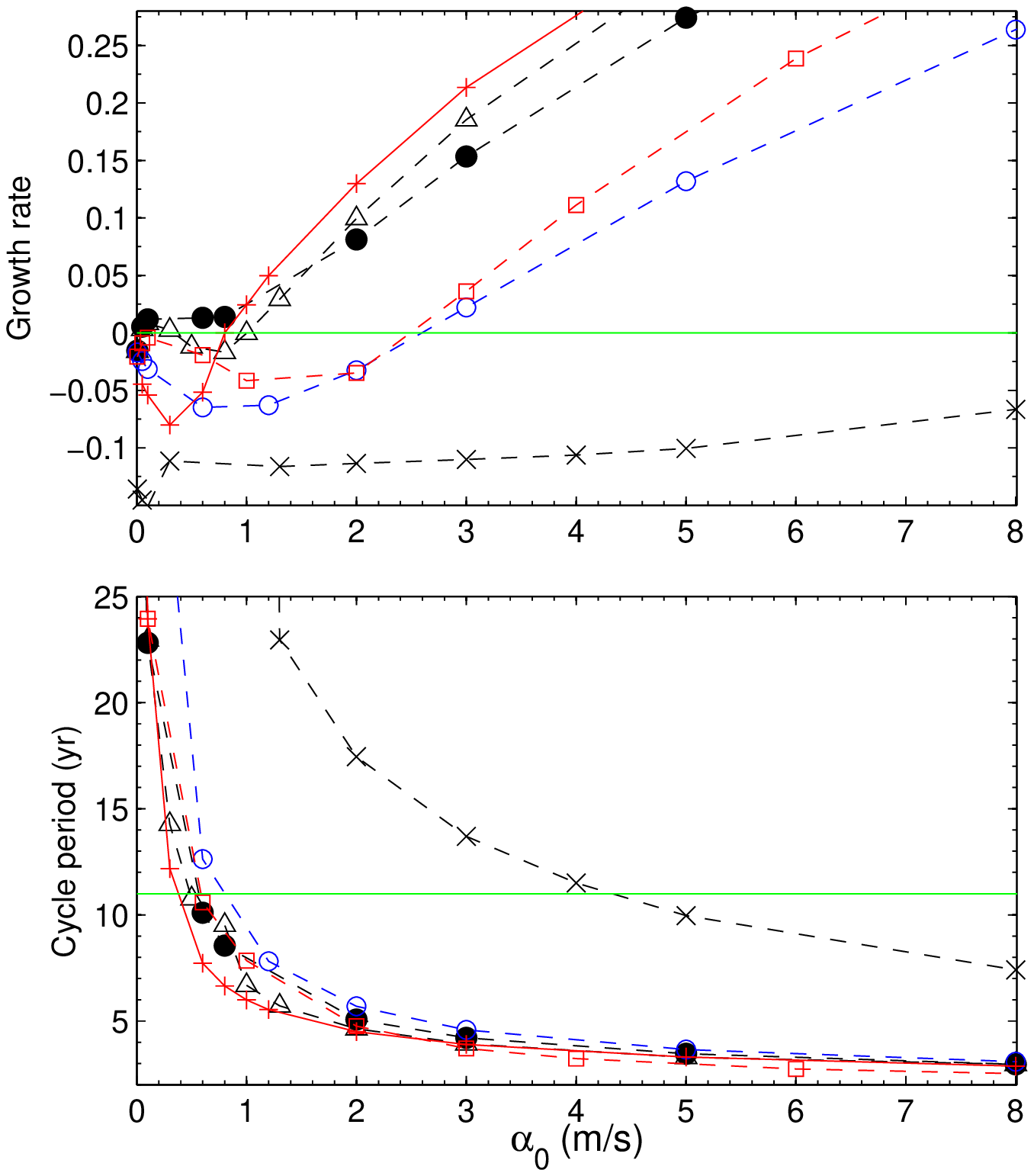}
\caption{Growth rates of toroidal flux (top) and the cycle periods (bottom) 
as functions of $\alpha_0$
from simulations in which the \bl\ source is related to the toroidal field in the 
tachocline.
Crosses represent simulations with \vp\ = 0 and $\etab = 5\times10^{11}$~\cmss. 
while for all others \vp\ = 35~\mps.
For filled points and squares $\etab = 5\times10^{11}$~\cmss\
and $7.5\times10^{11}$~\cmss, respectively.
Open circles: same as filled points but 
the {\it equatorward component} of meridional flow is zero,
triangles: same as filled points but
no radial shear in the NSSL
and red pluses: same as filled points except no \mc. 
}\label{grOmobstacal}
\end{figure}

%In our all previous simulations performed so far, the \bl\ source,
In all previous simulations presented above, the \bl\ source,
$\mean{B}$ in \Eq{BLsource}
was taken as the mean toroidal field over the whole CZ.
Now we performed some simulations taking the \bl\ source
$\mean{B}$ to be the mean toroidal field in the tachocline.
This is consistent with the prescription of the \bl\ process in
the traditional FTD model \citep{DC99,Kar14a}.
\Fig{grOmobstacal} displays the growth rates and the corresponding cycle periods
as functions of $\alpha_0$ from these simulations.
When there is no magnetic pumping (crosses), the dynamo is always decaying unless $\alpha_0$
is increased to a sufficiently larger value---consistent with the previous analysis.
For \vp\ = 35~\mps\ and $\etab = 5\times10^{11}$~\cmss\ (filled points), 
we observe growing solutions
for any values of $\alpha_0$ as long as it is above $\approx 0.02$~\mps.
For larger $\alpha_0$ we get shorter period---consistent with previous
simulations.
However for $\etab = 7.5\times10^{11}$~\cmss\ (squares) we get positive growth
only for $\alpha_0 \ge 3$~\mps\ but the cycle period becomes somewhat short.
Similar problem appears when either 
the equatorward component of
\mc\ is set to zero (open circles) or the full \mc\ 
is set to zero (red pluses).
For the case with no radial shear in the near-surface layer (triangles),
we get growing oscillations for $\alpha_0>1$~\mps\ but cycle periods 
are again shorter than 11 years.

\begin{figure}
\centering
\hspace*{-0.3cm}
\includegraphics[width=1.18\columnwidth]{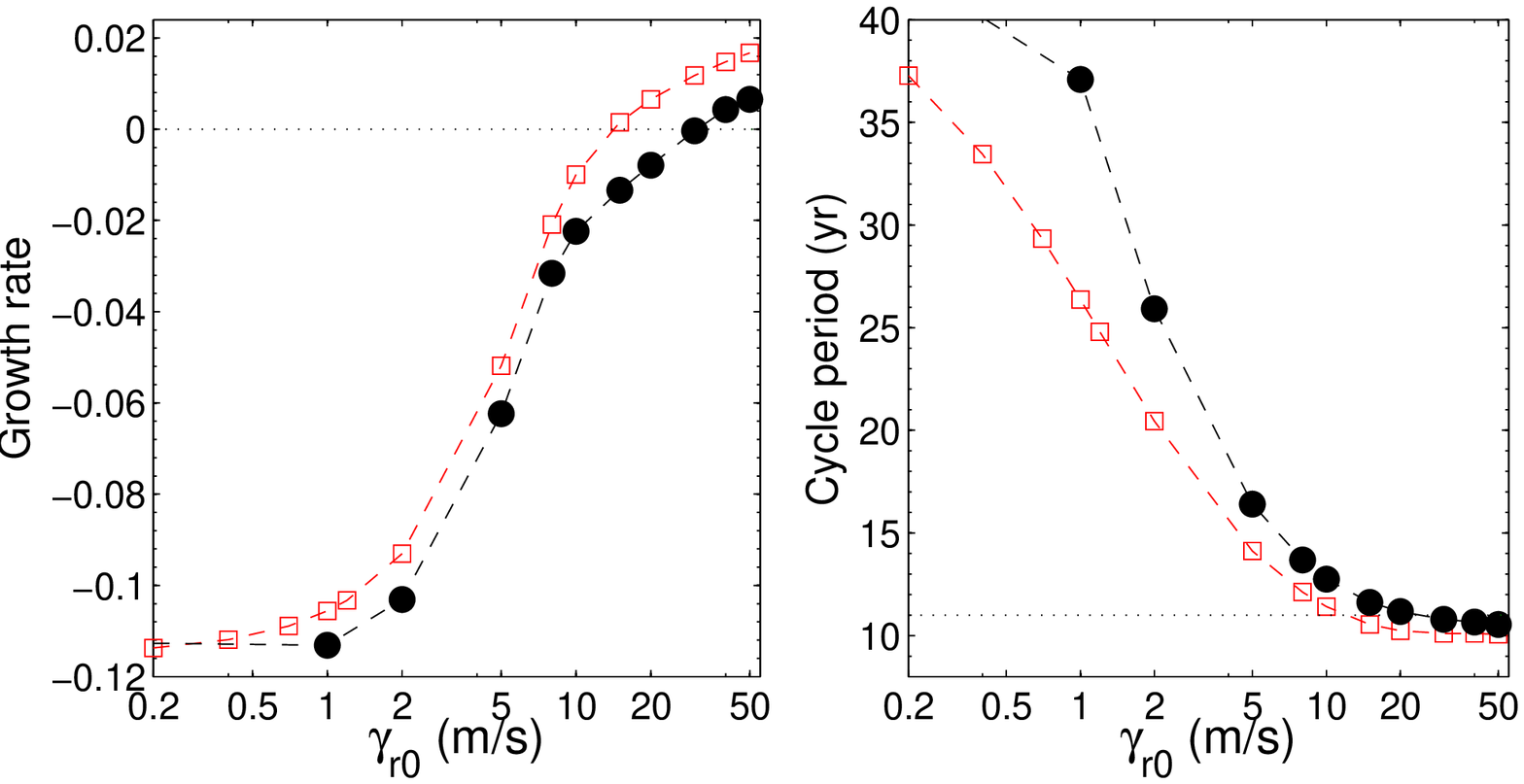}
\caption{
Growth rates (left panel) and cycle periods (right)
as functions of pumping speed \vp\ 
\blue{
(in log scale)
} 
for fixed values of other parameters.
The black points correspond to simulations in which the \bl\ source is
related to the mean toroidal field in the CZ and $\alpha_0 = 1.5$~\mps,
while for red squares, it is related to the tachocline toroidal field and $\alpha_0 = 0.6$~\mps.
For both sets $\etab = 5\times10^{11}$~\cmss\ and $\etas = 3\times10^{12}$~\cmss.
}
\label{varyp}
\end{figure}

\begin{figure}
\centering
\hspace*{-0.3cm}
\includegraphics[width=1.17\columnwidth]{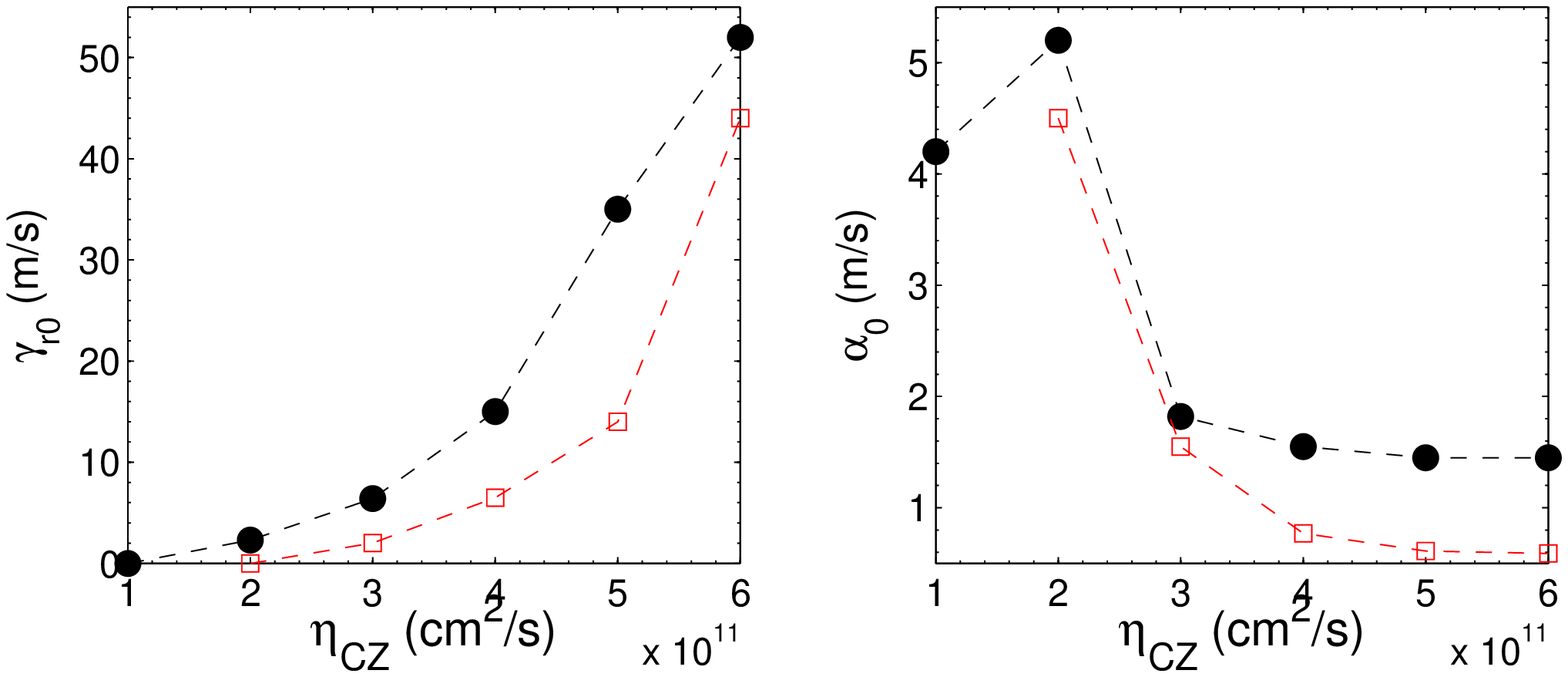}
\caption{Left panel: downward pumping speed \vp\ needed to get 
11-year cycle for different values of 
the bulk diffusivities $\etab$ at a fixed value of the 
surface diffusivity $\etas$ of $3\times10^{12}$~\cmss.
Black points and red squares represent two sets of simulations 
in which the \bl\ source is 
related to the mean toroidal field
in the CZ and in the tachocline, respectively.
Right panel: corresponding values of critical $\alpha_0$ needed to 
get stable solutions.
}
\label{varyep}
\end{figure}

One may wonder how sensitive we are to the choice of 35~\mps\  for the magnetic pumping.
To address this we take model with parameters: $\etab = 5\times10^{11}$~\cmss\
and $\alpha_0 = 0.6$~\mps\ (represented by a filled point in \Fig{grOmobstacal})
and perform a set of simulations by varying the pumping speed \vp\ only.
Growth rates and cycle periods from these simulations are displayed in \Fig{varyp}.
We observe that the dynamo growth increases rapidly with the increase of pumping
at lower values. This is expected because the 
downward pumping
suppresses the diffusion of the field across the solar surface as 
explained in \Sec{sec:decay}.
Following \citet{Ca12}, the dynamo growth rate becomes positive 
when the advection time by the pumping
is at least five times the diffusion time. 
Thus by setting, 
$5 d_{\rm{SL}}/\gamma_{r0} = d_{\rm{SL}}^2/\etas$,
(where $d_{\rm{SL}}$ is the depth of the surface pumping layer $= 0.1R$)
we obtain a rough estimate of the minimum \vp\ for the growing dynamo to be 20~\mps.
This is indeed seen in \Fig{varyp}. However above this value,
the pumping cannot help much and the dynamo growth will be limited by the 
cross-equator diffusion.
For other set of simulations in which the \bl\ source is related to the
mean toroidal field in CZ (black points in \Fig{varyp})
we observe a similar trend although slightly smaller growth rates.

In the right panel of \Fig{varyp}, we observe a rapid 
decrease of cycle period with the increase of
\vp\ at first and then saturation at larger values of \vp.
The reduction of cycle period 
is due to the fact that pumping quickly transports the field from surface to the deeper CZ. 
For \vp\ in the range [0.2--20]~\mps, the cycle period $\propto \gamma_{r0}^{-0.3}$,
which is a mild dependence compared to Eq.\ (5) of \citet{GD08} in the range [0.2--1.2]~\mps.
The discrepancy is due to the fact that we are using pumping only near the surface and 
higher diffusivity.

Now we address the question, what is the minimum value of pumping required 
to get an 11-year solar cycle for a given value of diffusivity in the CZ.
In \Fig{varyep}, we show this minimum pumping \vp\ for each value of 
$\etab$ that can produce
11-year cycle period.
We stress that in all these simulations $\etas$ and all other parameters, 
except $\alpha_0$ remain unchanged. The $\alpha_0$ is needed to vary 
because we make all the solutions critically/nearly stable
(with dynamo growth rate essentially very small and positive)---thus for this set of simulations $\alpha_0$ is not arbitrarily changed.
Red squares in \Fig{varyep} are obtained from
simulations in which the \bl\ source is related to the
tachocline toroidal field.
We observe that when $\etab$ is around $10^{11}$~\cmss\ or less, we do not need any magnetic pumping to get an 11-year cycle.
So at this point, our model essentially converges to the previous \ftdm.
We recall that all previous \ftdms\ need around this much diffusivity to get 
an 11-year cycle period \citep[e.g.,][]{Dik02,MNM11,JCSI13}.
The exact value of this critical diffusivity varies slightly from model to model as there are many other parameters that are not same
in all previous models. However, from our study we show that as we increase the diffusivity, we need to increase the downward pumping.
But when $\etab$ is above around $6\times10^{11}$~\cmss,
and the differential rotation is taken from helioseismic data,
we fail to produce 11-year cycle even by increasing the pumping to an 
arbitrarily large value.

The black points in \Fig{varyep} represent the results from simulations
in which the \bl\ source is related to the mean toroidal field in the CZ.
We observe that the results are essentially similar except that we need significantly larger pumping for the same parameters.

From the right panel of \Fig{varyep}, we note that the critical $\alpha_0$ needed to get stable dynamo cycle decreases with the increase 
of $\etab$. This counterintuitive behavior is found because at higher $\etab$, higher pumping is needed and this higher pumping 
reduces the diffusion of the field to make the dynamo easier.

%-----Variations of meridional flow ----
\subsection{Solar cycle with a shallow meridional circulation}
\label{sec:mc1}
\begin{figure}[t]
\hspace{-0.8cm}
\includegraphics[width=1.20\columnwidth]{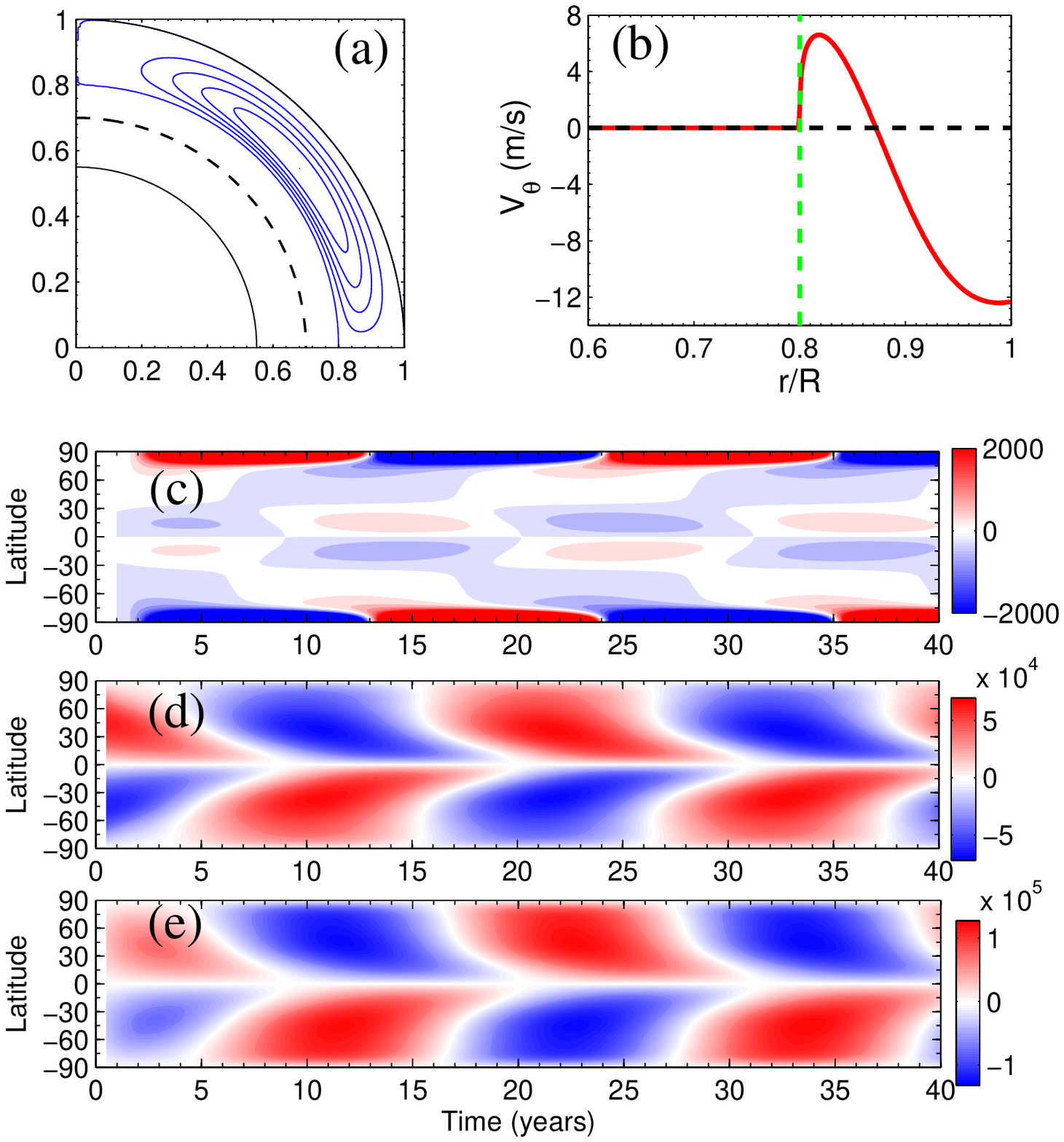}
\vspace{-1.1cm}
\caption{(a) Streamlines of a shallow meridional circulation 
with no flow below 
shown for the northern hemisphere only.
(b) Radial variation of $v_\theta$ at $45^\circ$ latitude.
Butterfly diagrams of: (c) surface radial field, (d) the mean 
toroidal field in the CZ and (e) the toroidal field in the tachocline.
In this simulation $\etab=5\times10^{11}$~\cmss\ and \vp\ = 9.8~\mps.
}
\label{fig:MCc1}
\end{figure}

We have understood that the downward pumping enhances the equatorward migration
produced in combination with the negative radial shear and 
the positive $\alpha^{\rm BL}$.
A clear equatorward migration is still observed when there is no
equatorward return flow in the deeper layer of CZ (\Figs{butOmannoeqMC}{butOmannoeqMCcs2}). 
This motivates us
to perform a simulation with a shallow \mc\footnote{To produce 
such a shallow meridional circulation,
we used the following modified parameters: 
$\beta_1 = 3.5$, 
$\beta_2 = 3.3$, $\Gamma = 3.4~$m,
$R_p = 0.8 R$, and the prefactor $r-R_p$ in \Eq{eqmc} is replaced by 
$1.0707 (r-R_p)^{0.3} [1-\mathrm{erf}\{(r-0.87R)/1.5\}]$.}, for example
residing only above $0.8R$
and no flow underneath. 
This scenario is very similar to the cases presented in
Figure~1 of \citet{HKC14} 
and Figure~5 of \citet{GD08}.
The former authors showed that this type of a shallow \mc\
is not able to produce an equatorward migration of the toroidal field
rather it produces a poleward migration, while 
the latter 
authors
showed that both an equatorward magnetic pumping and the dynamo wave due to 
the negative near-surface shear can produce an equatorward migration of the toroidal field, 
although the equatorward pumping alone produces the best result (see their Fig.\ 5).
However when we perform the simulation with this type of \mf,
we get a clear equatorward migration of toroidal field at low latitudes 
purely due to the negative near-surface shear
as shown in \Fig{fig:MCc1}.
Again this migration persists all the way to the tachocline as
shown in \Fig{fig:MCc1}(e).
This migration is now consistent with the observation.

The major difference between the present model and the \citet{HKC14} is that
they did not include radial pumping, the near-surface shear 
and the radial magnetic boundary condition at the surface.
On the other hand, \citet{GD08} used both radial and latitudinal pumpings
as well as a negative shear at low latitudes. Both the latitudinal pumping
and the negative shear were individually able to produce equatorward migration
(see their Fig.\ 5-7).
For our model to work we just need $9.8$~\mps\
downward magnetic pumping in the upper layer 
of the Sun and a radial magnetic boundary condition at the surface. 
Our value of pumping is much higher than \citet{GD08} used because
in our model, the turbulent diffusivity, particularly in the NSSL, is much higher.

In the present simulation, the \bl\ source is related to the toroidal field
in the tachocline. However, we get a similar result with a clear equatorward migration
when we perform the same simulation with the \bl\ source relating to 
the mean toroidal field in the CZ, except in this case 
we need \vp\ $= 11.5$~\mps. 
This reveals that, although the amount of pumping needed in each model 
varies depending on the diffusivity and other parameters, the qualitative idea works.
We recall that in earlier simulations, represented by open circles 
in \Figs{fig:grOmobs}{grOmobstacal}, when the equatorward component of the usual deep \mc, 
%that was extended from surface to the bottom of the CZ,
was set to zero, produced decaying solutions. However here for a shallow meridional flow with 
no flow underneath $0.8R$,
we get stable dynamo solution. The reason is that here we have an equatorward flow that helps to transport 
the toroidal field from high to low latitudes where the $\alpha^{\rm BL}$ works.

\section{Conclusion and Discussion}
\label{sec:conc}

We have revisited two fundamental features of the sunspot cycle: the equatorward
migration and the 11-year periodicity, using a dynamo model in which
the poloidal field is generated through the \bl\ process near the solar surface 
at low latitudes, whereas the toroidal field is generated through the stretching 
of poloidal field by differential rotation which is constrained by the
helioseismology. Our model also includes a single-cell meridional flow of which the surface flow 
is consistent with 
observations. The poleward flow near the surface is crucial for advecting the poloidal field 
towards the pole and thereby producing the observed poleward migration of the radial field 
on the surface. The poleward flow is also crucial to store the poloidal field in high latitudes for a sufficiently long time.
Whereas the equatorward component of the meridional flow gives the equatorward migration
of toroidal field near the bottom of CZ, consistent with previous \ftdms\ \citep{CSD95,Dur95}.
However, we show that equatorward \mf\ is not the only solution for this. We show that when we have a reasonable amount of magnetic pumping
in the upper layer of the Sun, the equatorward migration of the toroidal field is observed at low latitudes 
as a consequence of the dynamo wave (following Parker--Yoshimura sign rule with
positive $\alpha^{\rm BL}$ in the northern hemisphere and negative shear) 
even when there is no equatorward flow near the bottom of the CZ, 
although the cycle period becomes short (6 years rather than 11 years) when we use the
helioseismic differential rotation profile.
By reducing the surface diffusion, $\etas$ and $\alpha_0$, we can 
achieve an 11-year period in this case too,
but the dynamo tends to produce a quadrupolar field which is not solar-like
\cite[see][who demonstrated that a strong surface diffusion is crucial 
to get dipolar fields]{CNC04}. 
We note that if the surface poleward flow is also switched off in the dynamo wave case,
then the cycle period becomes short again and the dynamo becomes weaker 
\citep[consistent with the ideas of][]{RS72}.

In the case where we use the fully observed differential rotation
we get a FTD type solution modified by the NSSL
 consistent with \citet{GD08} and the distributed model of \citet{KKT06}.
Because both terms play a role, we get a clear equatorward migration with a
shallow \mf\ residing above $0.8R$ (\Fig{fig:MCc1}). 
The main difference between dynamo wave and FTD solutions is in the value of 
$\alpha_0$ required to get growing solutions with the correct period. 

In all these simulations, the magnetic pumping plays a crucial role in suppressing the diffusion
of fields across the surface. This helps us to achieve dynamo at a significantly
higher value of diffusivity in the bulk. For pumping \vp\ = 35~\mps\ and surface diffusivity
$\etas=3\times10^{12}$~\cmss, we find a dynamo solution with an 11-year period 
at a bulk diffusivity $\etab=5\times10^{11}$~\cmss, which is
5 times larger than the allowed value when there is no magnetic pumping 
(see \Fig{varyep} for details).
However, if the Babcock-Leighton source is related to the toroidal field
near the bottom of the CZ, which is the usual prescription in flux transport 
dynamo models, then our model works even at smaller magnetic pumping 
(about 10~\mps\ for above parameters).

%For $\etab$ larger than $\etab=5\times10^{11}$~\cmss, in the part of parameter space we have sampled, we do not get an 11-year cycle. 
For $\etab > 5\times10^{11}$~\cmss, in the part of parameter space we have sampled, we do not get an 11-year cycle. 
We could obtain dynamo solutions with the correct period and equatorial
propagation at a diffusivity of $10^{12}$~\cmss\ using a simplified 
version of the differential rotation, which gives an indication of
our sensitivity to the differential rotation, pumping and choice of diffusivity profile. 
While this is still an order of magnitude
less than expected from mixing length arguments, it is close to the values suggested from the study of \citet{CS16}.
All previous \ftdms\ were constructed with a diffusivity much weaker than $10^{12}$~\cmss\ \citep{MNM11,MT16}. 
Although Choudhuri and his colleagues \cite[][and the publications later]{CNC04} have used the diffusivity for poloidal field
$\approx10^{12}$~\cmss, the diffusivity for toroidal field is reduced by a factor of about 60. 
On the other hand, \citet{KO12} have used a diffusivity $\sim 10^{13}$~\cmss\ in the bulk of CZ, consistent with the mixing-length value, 
but they reduce it by four orders of magnitudes below $0.75R$
and a diamagnetic pumping is considered there.
In this article, we have shown that a moderately high value of the diffusivity
($5\times 10^{11}$ to $10^{12}$ \cmss) is plausible 
given a sufficiently strong pumping near the surface. 
Such pumping is possible near the solar surface and 
has been shown to be necessary in
order to make the FTD model compatible with surface observations \citep{Ca12}.

\bibliographystyle{apj}
\begin{acknowledgements}
We thank Mark Miesch, Jie Jiang and the anonymous referee for 
careful review this article and providing valuable 
comments which improved the clarity of the presentation.
BBK is supported by the NASA Living With a Star Jack Eddy Postdoctoral Fellowship Program,
administered by the University Corporation for Atmospheric Research.
The National Center for Atmospheric Research is sponsored by the National Science Foundation.
\end{acknowledgements}

\bibliography{paper}
\end{document}